\documentclass[useAMS,usenatbib]{mn2e}

\usepackage{graphicx}
\usepackage{amsmath}
\usepackage{graphicx}
\usepackage[T1]{fontenc}
\usepackage[utf8]{inputenc}
\usepackage{mathtools}  
\usepackage{amssymb}
\usepackage{tabulary}
\usepackage{booktabs}

\newcommand{\Rd}{\textup{\textrm{\textcolor{black}{d}}}}
\newcommand{\ee}{\mathrm{e}}
\newcommand{\ii}{\mathrm{i}}
\newcommand{\JVLA}{\textcolor{black}{\text{VLA}}}
\usepackage[usenames,dvipsnames,svgnames,table]{xcolor}
\newcommand{\ATM}[1]{\textcolor{black}{{#1}}}
\newcommand{\ATMNEW}[1]{\textcolor{black}{{#1}}}
\newcommand{\Bessel}{\textcolor{black}{\text{Bessel}}}
\newcommand{\Sincc}{\textcolor{black}{\text{sinc}}}
\title[BDWFs for data compression and FoI shaping]{Using baseline-dependent window functions 
for data compression and field-of-interest shaping in radio interferometry}
\author[M. T. Atemkeng, O. M. Smirnov, C. Tasse, G. Foster and J. Jonas]{M. T. 
Atemkeng$^{1}$\thanks{E-mail: m.atemkeng@gmail.com}, O. M. Smirnov$^{12}$, C. Tasse$^{31}$, G. Foster$^{12}$, J. Jonas$^{12}$ \\
$^1$Department of Physics and Electronics, Rhodes University, PO Box 94, Grahamstown, 6140, South Africa\\
$^2$SKA South Africa, 3rd Floor, The Park, Park Road, Pinelands, 7405, South Africa\\
$^3$GEPI, Observatoire de Paris, CNRS, Universite Paris Diderot, 5 place Jules Janssen, 92190 Meudon, France}
\begin{document}

\date{Accepted 2016 July 07. Received 2016 June 09; in original form 2016 March 09.}

\pagerange{\pageref{firstpage}--\pageref{lastpage}} \pubyear{2016}

\maketitle

\label{firstpage}

\begin{abstract}

In radio interferometry, observed visibilities are intrinsically sampled at some interval in time and frequency. 
Modern interferometers are capable of producing data at very high time and frequency resolution; practical 
limits on storage and computation costs require that some form of data compression be imposed. The 
traditional form of compression is a simple averaging of the visibilities over coarser time and frequency bins. This
has an undesired side effect: the resulting averaged visibilities ``decorrelate'', and do so differently depending 
on the baseline length and averaging interval. This translates into a non-trivial signature in the image domain 
known as ``smearing'', which manifests itself as an attenuation in amplitude towards off-centre sources. 
With the increasing fields of view and/or longer baselines employed in modern and future instruments, the trade-off between
data rate and smearing becomes increasingly unfavourable. In this work we investigate alternative approaches to low-loss data
compression. We show that averaging of the visibility data 
can be treated as a form of convolution by a boxcar-like window function, and that by employing alternative 
baseline-dependent window functions a more optimal interferometer smearing response may be induced. In particular, 
we show improved amplitude response over a chosen field of interest, and better attenuation of sources outside the field of interest. The main cost of this technique is a reduction in nominal sensitivity; we investigate the smearing vs. 
sensitivity trade-off, and show that in certain regimes a favourable compromise can be achieved. We show the application of this technique to simulated data from the \ATM{Karl G. Jansky Very Large Array (\JVLA) and the European Very-long-baseline interferometry Network (EVN)}.
\end{abstract}

\section[]{Introduction}
A radio interferometer measures complex quantities called \emph{visibilities}, which, following the van Cittert-Zernike 
relation~\citep{thompson1999fundamentals,thompson2001fundamentals}, correspond to Fourier modes of the sky brightness distribution, corrupted by various instrumental 
and atmospheric effects. 
One particular effect, known as \emph{time} and \emph{bandwidth decorrelation} (or its equivalent in the image plane referred as smearing) occurs 
when the visibilities are averaged over a time and frequency bin of non-zero extent \citep{bridle1989wide,bridle1999bandwidth}. This unavoidably happens in the correlator 
(since the correlator output is, by definition, an average measurement over some interval), and also if data is further 
averaged in post-correlation for the purposes of compression and to reduce computational cost.

The effect of smearing is mainly a decrease in the amplitude of off-axis sources. This is easy to understand: the visibility contribution of a point source of flux $S$ located in the direction given by the unit vector $\bmath{\sigma}$ is given by
\begin{equation}
V = S \exp \big \{  \frac{2\pi i}{\lambda} \bmath{u}\cdot(\bmath{\sigma}-\bmath{\sigma}_0) \big \},
\end{equation}
where $\bmath{u}$ is the baseline vector (in a coordinate system fixed to the sky), 
and $\bmath{\sigma}_0$ is the phase centre (or fringe stopping centre) of the observation.
The complex phase term above rotates as a function of frequency (due to the inverse scaling with $\lambda$) and time (due to
the fact that $\bmath{u}$ changes with time, at least in an Earth- or orbit-based interferometer). 
Taking a rotating complex vector average over a time/frequency bin then results  in a net loss of amplitude. The effect increases 
with baseline length and distance from phase centre. Besides reducing apparent source flux, smearing also distorts the 
point spread function (PSF), since different baselines (and thus different Fourier modes) are attenuated differently.
% \textcolor{red}{Reviewer comments: 4) You should add something about previous work in this field: mention e.g.
%     Thompson, Moran section 6.3 "Effect of Bandwidth" where a Gaussian passband
%     is mentioned.}
\ATM{The issue of time-frequency averaging has been addressed in the past by among others~\citet[]{thompson2008interferometry}, where a Gaussian taper has been used to eliminate smearing at the edges of the field of view (FoV). However, the problem of eliminating smearing to about 1\% or less within the FoV while compressing the data to an acceptable level has not been satisfactorily addressed before.}

Throughout this work we use the term Field of Interest (FoI) which we
differentiate from the FoV. For this work the FoV is related to
angular scale of the primary beam where as the FoI is a parameter of the 
scientific observation case, which may be related to the size of the primary
beam but is not a necessary requirement. Thus, the FoI can be an adjustable
parameter in the window functions we present.

In the era of large interferometers, where computation (and thus data size)
becomes one of the main cost drivers, it is in principle desirable to average
the data down as much as possible, without compromising the science goals.
There are natural limits to this: firstly, we still need to critically sample
the $uv$-plane, secondly, we need to retain sufficient spectral resolution,
thirdly, we do not want to average (at least pre-calibration) beyond the
natural variation of the calibration parameters, and fourthly, we want to keep
smearing at acceptable levels in order not to lose too much signal. In this
work, we concentrate specifically on the decorrelation/smearing problem. Typical
observation cases are:
\begin{itemize}
\item Surveys where it is desirable to have a flat response across the largest
possible FoI. This can be achieved by facet imaging but at a computational cost.
By increasing the size of the facets a better balance between computational cost
and image response can be achieved.
\item Deep imaging where the it is useful to suppress bright sources in the
primary beam sidelobes outside of the FoI. Even with smearing of sources far
from the phase centre bright sources can contribute significant flux and PSF
sidelobe artefacts to the image.
\item Very Long Baseline Interferometry (VLBI), decorrelation is more severe, so
the effective FoI is determined by the smallest time/frequency bin size that a
correlator can support, and is normally much smaller than the primary beam
\citep{keipema2015sfxc}. Modern VLBI correlators overcome this by employing a
technique where the signal is correlated relative to multiple phase centres
simultaneously, thus effectively ``tiling'' the primary beam by multiple FoIs.
This has a computational cost that scales linearly with the number of phase
centres.
\end{itemize}
Smearing can be seen to be a useful side effect, anything outside the desired
FoI (by definition) is unwanted signal.
The primary beam pattern of any real antenna features sidelobes and backlobes 
that extend across the entire sky, albeit at a relatively faint level. The faintness makes sidelobes useless for imaging any 
but the brightest sources. However, the 
sum total signal from all the sources in the primary beam sidelobes, modulated by their PSF sidelobes, contributes an unwanted global 
background called the \emph{far sidelobe confusion noise} (FSCN), in very deep observations this may in principle become a
bottleneck \citep{icea-fscn}. In other cases, individual extremely bright
radio sources such as Cygnus A or 
Cassiopeia A can contribute confusing signal from even the most distant  sidelobe: the LOFAR telescope \citep{LOFAR} has to deal 
with these so-called ``A-team'' sources 
on a routine basis. By suppressing distant off-axis sources, smearing somewhat alleviates both the FSCN and A-team problems.

When considering a short sequence of visibilities measured on one baseline, we can consider averaging as a convolution of the 
true visibility by a boxcar function corresponding to the $uv$-extent of the averaging bin, followed by sampling at the 
centre of each bin. Convolution in the visibility plane corresponds to multiplication of the image by an 
\emph{image plane response function} 
that is the Fourier transform  of the convolution kernel i.e the window function; 
the Fourier transform of a boxcar is a \Sincc-type taper.

If we consider the entire $uv$-plane, averaging is only a pseudo-convolution, since the different $uv$-bins (and thus
their boxcars) will have different sizes and shapes as determined by baseline length and orientation. Still, we can 
qualitatively view smearing  as some kind of cumulative effect of an ensemble of image-plane tapers corresponding to all the 
different boxcars\footnote{For completeness, we should note that  this ``smearing taper'' is not the only tapering effect 
at work in interferometric imaging. Firstly, antennas have a non-zero 
physical extent: a measured visibility is already convolved by the aperture illumination functions  of each pair of 
antennas. The resulting image-plane taper is exactly what the primary beam is. Secondly, most imaging software employs 
convolutional gridding followed by an fast Fourier transform, which produces an additional taper that suppresses aliasing of sources from 
outside the imaged region.}. 

What if we were to employ weighted averaging instead of simple averaging (whether in the correlator, or in post-processing)? 
This would correspond to a  pseudo-convolution of the $uv$-plane by some ensemble of \emph{window functions}, 
different from boxcars, which would obviously yield different image-plane tapers, and thus result in different 
smearing response. Filter theory suggests that a window function can be tuned to achieve some desired tapering response. 
An optimal taper would be one that was maximal across the desired FoI, and minimal outside it. In this work, 
we apply filter theory to derive a set of baseline-dependent window functions (BDWFs)
%\textcolor{red}{Reviewer comment: Notation \& Nomenclature: 1)}
\ATM{\footnote{\ATM{BDWFs are functions of a baseline's length and orientation; the East-West component rotates as a function of time, while  the South-North does not. For the same length of time,  two equal} \ATM{length baselines with different orientations will sweep  unequal $uv$-space and therefore result in a different degree of decorrelation.}
}}
that approximate this more optimal smearing 
behaviour. The trade-off is an increase in thermal noise, since minimum noise can only be achieved with 
unweighted averaging. We show that this effect can be partially mitigated through the use of \emph{overlapping window functions}. 
\citet{offringa-filtering} have investigated a similar approach in the context of suppressing signals towards specific off-axis
sources. 

In the era of the Square Kilometre Array (SKA) and its pathfinders, where dealing with the huge data volumes is one of
the main challenges, use of BDWFs potentially offers additional leverage in optimising radio observations. 
Decreased smearing across the FoI allows for more aggressive data averaging, thus reducing storage and compute costs. 
The trade-off is a loss of sensitivity, which pushes up observational time requirements. However, the decrease in smearing
and noise from A-team sources could, conceivably, make up for some of the nominal sensitivity loss. 
In the VLBI case, use of BDWFs potentially offers an increase in effective FoI at a given correlator dump rate, or 
equivalently, the ability to tile the primary beam with fewer phase centres, allowing
for smaller correlators.

\section{Overview and problem statement}
\newcommand{\VV}{\mathcal{V}}
\newcommand{\PP}{\mathcal{P}}
\newcommand{\VVM}{\textcolor{black}{\widehat{\mathcal{V}}}}%^\mathrm{M}}
\newcommand{\WW}{\mathcal{W}}
\newcommand{\II}{\mathcal{I}}
\newcommand{\IID}{\mathcal{I}^\mathrm{D}}
\newcommand{\IIDI}{\mathcal{I}^\mathrm{DI}}
\newcommand{\EE}{\mathcal{E}}
\newcommand{\FF}{\mathcal{F}}
\newcommand{\HH}{\mathcal{H}}
\newcommand{\TT}{\mathcal{T}}
\newcommand{\NN}{\mathcal{N}}
\newcommand{\uu}{\bmath{u}}
\newcommand{\Btf}{\mathsf{B}^{[\Delta t\Delta\nu]}}
\newcommand{\Babtf}{\mathsf{B}^{[\alpha\Delta t,\beta\Delta\nu]}}
\newcommand{\Bab}{\mathsf{B}^{[\alpha\beta]}}
\newcommand{\Buv}{\mathsf{B}^{[uv]}}
\newcommand{\Bij}{\mathsf{B}}
\newcommand{\Ptf}{\Pi^{[t\nu]}}
\newcommand{\Puv}{\Pi^{[uv]}}
\newcommand{\Vm}{\textcolor{black}{\widehat{V}}}%^\mathrm{M}}
\newcommand{\Vs}{V^\mathrm{S}}

The following formalism deals with visibilities both as functions (i.e. entire distributions on the $uv$-plane), 
and single visibilities (i.e. values of those functions at a specific point). To avoid confusion between functions in
functional notation and their values, we will use $\VV$ or 
$\VV(u,v)$ to denote functions, and $V$ to denote individual visibilities. Likewise, $\II(l,m)$ denotes a function 
on the $lm$-plane i.e. an image. The symbol $\delta$ always denotes the Kronecker delta-function.

Depending on whether we want to consider polarisation or not, $\VV$ can be taken to represent either 
scalar (complex) visibilities, or $2\times2$ complex visibility matrices  as per the radio interferometer 
measurement equation (RIME) formalism \citep{smirnov2011revisiting}. Likewise, $\II$ can be treated as a scalar 
(total intensity) image, or a $2\times2$ brightness matrix distribution. The derivations below 
are valid in either case.

We shall use the symbols $\mathbf{u}=(u,v)$ or $\mathbf{u}=(u,v,w)$ to represent baseline coordinates in units of wavelength.
%, and 
%$\mathbf{u}^\mathrm{m}$ for units of metres, with $\mathbf{u} = \mathbf{u}^\mathrm{m}/\lambda = \mathbf{u}^\mathrm{m}\nu/c$.

\subsection{Visibility and relation with the sky}
\label{sec:visSky}
An interferometer array measures the quantity $\VV(u,v,w)$, known as the visibility function.
Here, the coordinates $u,v$ and $w$ are vector components in units of wavelength, describing the distance between 
antennas $p$ and $q$, called the \emph{baseline}. The $w$ axis is oriented towards the \emph{phase centre} of the observed field,
while $u$ points east and $v$ north. Given a sky distribution $\II_0(l,m)$, where $l,m$ are the direction cosines,
the nominal observed visibility is given by the van Cittert-Zernike theorem \citep{thompson1999fundamentals,thompson2001fundamentals} as
\begin{equation}
\VV^\mathrm{nom}(u,v) =\iint\limits_{lm} \frac{\II_0(l,m)}{\sqrt{1-l^2 - m^2}}\,\ee^{-2\pi\ii\phi (u,v,w)}\Rd l\Rd m, \label{eq:visSky:nom}
\end{equation} 
where $\phi(u,v,w)=ul+vm+w(n-1)$, and $n=\sqrt{1-l^2 - m^2}$ (the $n-1$ term comes about when fringe 
stopping is in effect, i.e. when 
the correlator introduces a compensating delay to ensure $\phi=0$ at the centre of the field, otherwise the term is simply $n$). 

Given a pair of antennas $p$ and $q$ forming a baseline $\bmath{u}_{pq}=(u_{pq},v_{pq},w_{pq})$, 
and taking into account the \emph{primary beam} patterns $\EE_p(l,m)$ and $\EE_q(l,m)$ that define the directional sensitivity of 
the antennas, this becomes 
\begin{equation}
\VV_{pq}(u,v)=\iint\limits_{lm} \frac{\EE_p \II_0 \EE_q^H}{\sqrt{1-l^2 - m^2}}\,\ee^{-2\pi\ii\phi (u,v,w)}\Rd l\Rd m, \label{eq:visSky}
\end{equation}
where $^H$ represents the conjugate transpose. The first term being integrated is the \emph{apparent sky seen by baseline} $pq$,
\begin{equation}
\II_{pq} = \frac{\EE_p \II_0 \EE_q^H}{\sqrt{1-l^2 - m^2}},
\end{equation} 
which in general can be variable in time and frequency. 
For simplicity, let us assume that both the sky and the primary beam are constant (invariant in time and frequency), and that the primary beam is the same for all stations. All baselines will then see the same apparent sky throughout the measurement process. Let us designate this by $\II$. Assuming a small FoI ($n\to 1$) and/or a co-planar array ($w=0$), the above equation becomes a simple 2D Fourier transform :
\begin{equation}
\VV(u,v)=\iint\limits_{lm} \II\,\ee^{-2\pi\ii(ul+vm)}\Rd l \Rd m, \label{eq:visSky:2D}
\end{equation} 
or in functional form,
\begin{equation}
\VV=\FF\{\II\},~~\II=\FF^{-1}\{\VV\}.
\end{equation}
\textcolor{black}{we will} refer to $\VV$ as the \emph{ideal} visibility distribution (as opposed
to the \emph{measured} distribution, which is corrupted by averaging in the correlator, as \textcolor{black}{we will} explore below).

Note that the effect of the primary beam can alternatively be expressed in terms of a convolution with its Fourier transform, the \emph{aperture 
illumination function}  $\mathcal{A}_p(u,v)$. In functional form:
\begin{equation}
\VV_{pq} = \mathcal{A}_p \circ \VV^\mathrm{nom}_{pq} \circ \mathcal{A}_q^H.\label{eq:visSky:conv}
\end{equation} 

\subsection{Imaging, averaging and convolution}
\label{sec:AvgCon}

Earth rotation causes the baseline to rotate in time,
%which we can denote by $\bmath{u}^\mathrm{}_{pq}(t)$,
%$\bmath{u}^\mathrm{m}_{pq}=\bmath{u}^\mathrm{m}_{pq}(t)$. 
the baseline in units of wavelength 
can be treated as a function of frequency and time (from this point on-wards we shall assume that the sky is constant across the
range of frequencies being observed):
\begin{equation}
\label{eq:uvtf}
\bmath{u}_{pq}(t,\nu) = \bmath{u}^\mathrm{}_{pq}(t)\nu/c.
\end{equation} 
This, in turn, allows us to rewrite the visibility in eq.~(\ref{eq:visSky:2D}) as a per-baseline function of $t,\nu$:
\begin{equation}
V_{pq}(t,\nu)=\iint\limits_{lm} \II\,\ee^{-2\pi\ii(u_{pq}(t,\nu)l+v_{pq}(t,\nu)m)}\Rd l \Rd m. 
\label{eq:visSky:2Dtf}
\end{equation} 

Synthesis imaging recovers the so-called ``dirty image'': the inverse Fourier transform of the measured visibility distribution $\VVM$ sampled by a number of baselines $pq$ at discrete time/frequency points. Inverting the Fourier transform produces the dirty image:
\begin{equation}
\label{eq:imaging}
\IID = \FF^{-1}\{ \WW\cdot\VVM \} ,
\end{equation}
where $\WW$ is the (weighted) sampling function -- a ``bed-of-nails'' function that is non-zero at points where we 
are sampling a visibility, and zero elsewhere. If $\VVM=\VV$, then this can also be expressed as a convolution of the apparent sky by the \emph{point spread function}
$\PP$:
\begin{equation}
\IID = \PP\circ\II,~~\PP=\FF^{-1}\{\WW\}.
\end{equation}
Designating each baseline as $pq$, and each time/frequency point
as $t_k,\nu_l$, we can represent $\WW$ by a sum of  ``single-nail'' functions 
$\WW_{pqkl}$:
\begin{equation}
\WW = \sum_{pqkl} \WW_{pqkl} = \sum_{pqkl} W_{pqkl} \delta_{pqkl},
\end{equation}
where $\delta_{pqkl}$ is a delta-function shifted to the $uv$-point being sampled:
\begin{equation}
\delta_{pqkl}(\bmath{u}) = \delta(\bmath{u}-\bmath{u}_{pq}(t_k,\nu_l))
\end{equation}
and $W_{pqkl}$ is the 
associated weight. The Fourier transform being linear, we can rewrite eq.~(\ref{eq:imaging}) as 
\begin{equation}
\label{eq:imaging2}
\IID = \sum_{pqkl} W_{pqkl} \mathcal{F}^H\{ \VVM_{pqkl} \},
\end{equation}
where 
\begin{equation}
\label{eq:imaging2a}
\VVM_{pqkl} = \delta_{pqkl} \Vm_{pq}(t_k,\nu_l)
\end{equation}
i.e. the visibility distribution corresponding to the single visibility sample $pqkl$. We can further rewrite eq.~(\ref{eq:imaging})
again as
\begin{equation}
\label{eq:imaging3}
\IID =  \sum_{pqkl} W_{pqkl} \FF^{-1}\{ \VVM_{pqkl} \},
\end{equation}
which shows that the dirty image $\IID$ can be seen as a weighted sum of images corresponding to the individual visibility samples $pqkl$ (each such image essentially being a single fringe pattern).

In the ideal case, we would be measuring instantaneous visibility samples, and (assuming no other instrumental corruptions),
we would have $\VVM\equiv\VV$, with 
\begin{equation}
\Vm_{pq}(t_k,\nu_l)  = \VV(\bmath{u}_{pq}(t_k,\nu_l)),
\end{equation}
and consequently,
\begin{equation}
\label{eq:inst-sampled-visibility}
\VVM_{pqkl} = \delta_{pqkl} \VV,
\end{equation}
resulting in what \textcolor{black}{we will} call the \emph{ideal} dirty image 
$\IIDI$:
\begin{equation}
\label{eq:imaging:DI}
\IIDI =  \sum_{pqkl} W_{pqkl} \PP_{pqkl} \circ \II,~~\PP_{pqkl}=\FF^{-1}\{ \delta_{pqkl} \}
\end{equation}
That is, in the ideal case, each term in the weighted sum is equal to the apparent sky $\II$ convolved with a PSF representing a single visibility sample, $\PP_{pqkl}$.

However, an actual interferometer is necessarily non-ideal, in that it can only measure the average visibility over a some time-frequency bin given by
the \emph{time} and \emph{frequency sampling intervals} $\Delta t,\Delta \nu$, which \textcolor{black}{we will} call the \emph{sampling bin}
\begin{equation}
\Btf_{kl} = \bigg [ t_k-\frac{\Delta t}{2},t_k+\frac{\Delta t}{2} \bigg ]
\times
\bigg [ \nu_l-\frac{\Delta\nu}{2},\nu_l+\frac{\Delta\nu}{2} \bigg ],  \label{eq:chap3resamplingbin}
\end{equation}
This measurement can be represented by an integration:
\begin{equation}
\Vm_{pqkl} = \frac{1}{\Delta t \Delta \nu} 
\iint\limits_{\Btf_{kl}}
\VV(\bmath{u}_{pq}(t,\nu))\Rd\nu \Rd t.
\label{eq2:conti}
\end{equation}

\begin{figure*}
\includegraphics[width=0.75\columnwidth]{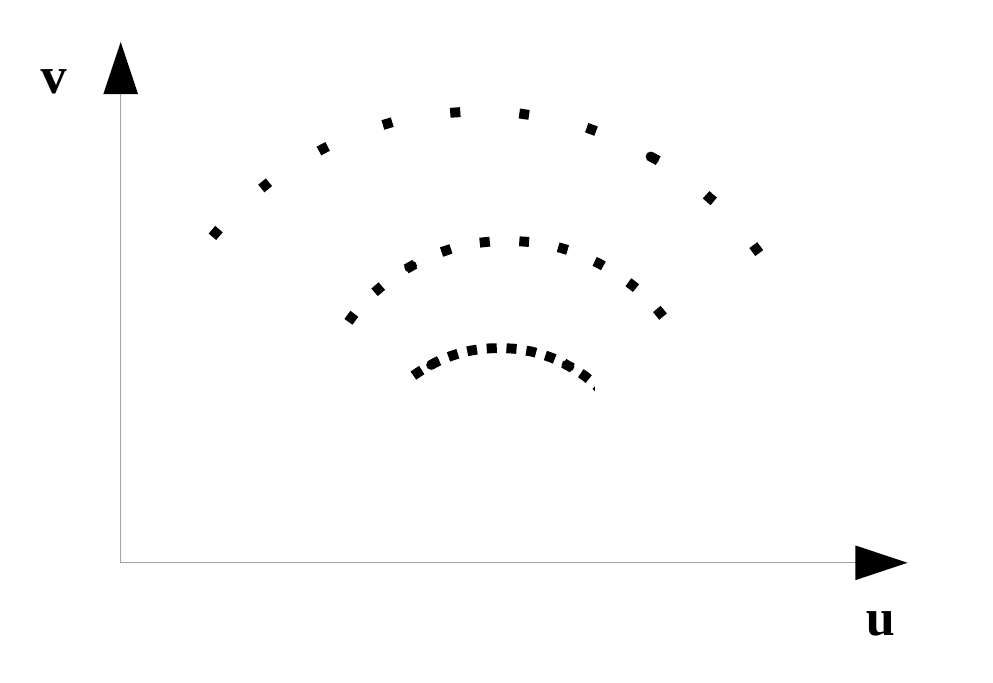}
\includegraphics[width=0.75\columnwidth]{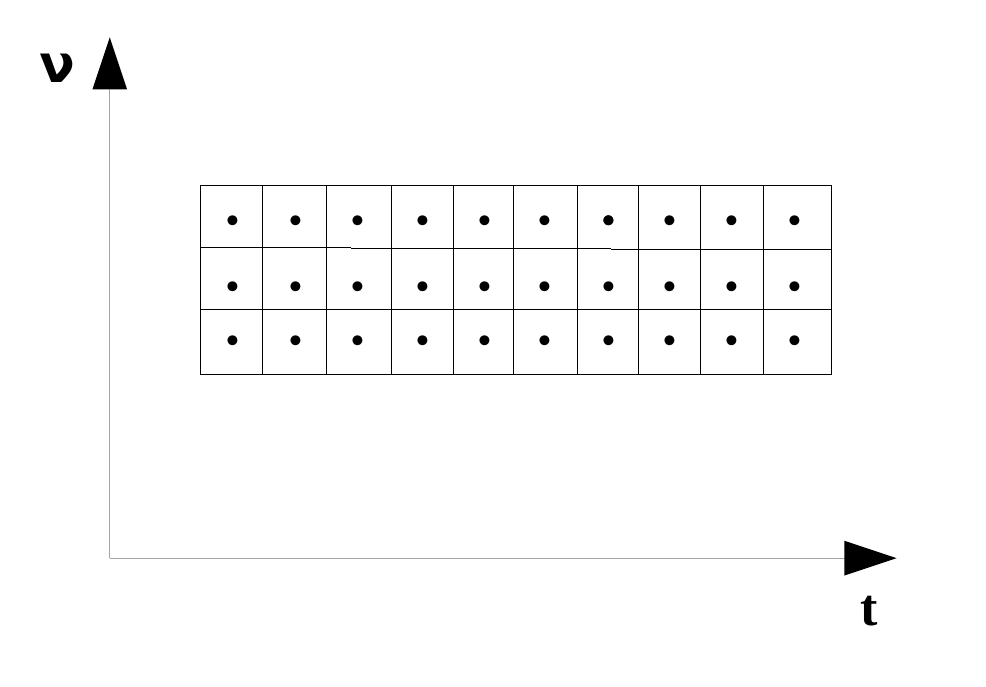}
\caption{Schematic of $uv$-coverage for 
regularly spaced time-frequency samples \ATM{(left: $uv$-space, right: $t\nu$-space).} Baselines with a longer East-West component sweep out longer tracks between successive 
integration's.}\label{fig:uvcov}
\end{figure*}

Inverting the relation of eq.~(\ref{eq:uvtf}), we can change variables to express this as an integration over the 
corresponding bin $\Buv_{pqkl}$ in $uv$-space:
\begin{equation}
\Vm_{pqkl} = \frac{1}{\Delta t \Delta \nu} 
\iint\limits_{\Buv_{pqkl}}
\VV_{pq}(u,v)\bigg| \frac{\partial(t,\nu)}{\partial(u,v)}\bigg| \Rd u \Rd v,
\label{eq2:conti:uv}
\end{equation}
where $\Buv_{pqkl}$ is the corresponding bin in $uv$-space. Note that the sampling bins in $t\nu$-space are
perfectly rectangular \ATM{(Fig.~\ref{fig:uvcov}, right)} and do not depend on baseline (assuming baseline-independent averaging), while the 
sampling bins in $uv$-space are elliptical arcs, and do depend on baseline (hence the extra $pq$ index). 
Assuming a bin small enough that the fringe rate $\partial\bmath{u}/\partial t$ is approximately constant over 
the bin, we then have \begin{equation}
\Vm_{pqkl} \sim \iint\limits_{\Buv_{pqkl}}
\VV (\bmath{u}) \Rd\bmath{u},
\label{eq2:conti:uv1}
\end{equation}

Now, let us introduce a \emph{normalised boxcar window function}, $\Ptf$ 
\begin{equation}
\Ptf(t,\nu) = \bigg \{ \begin{array}{cl}
\frac{1}{\Delta t\Delta\nu}, &  |t|\leq\Delta t/2,~~|\nu|\leq\Delta\nu/2 \\
0, & \mathrm{otherwise},
\end{array}
\end{equation}
using which we may re-write eq.~(\ref{eq2:conti}) as
\begin{equation}
\Vm_{pqkl} =  
\iint\limits_{\infty}
\VV_{pq}(t,\nu) \Ptf(t-t_k,\nu-\nu_l) \Rd t \Rd \nu,
\end{equation}
which can also be expressed as a convolution:
\begin{equation}
\Vm_{pqkl} = [ \VV_{pq} \circ \Ptf ](t_k,\nu_l),
\end{equation}
Likewise, eq.~(\ref{eq2:conti:uv}) can also be rewritten as a convolution in $uv$-space:
\begin{equation}
\Vm_{pqkl} = [ \VV_{pq} \circ \Puv_{pqkl} ](\bmath{u}_{pq}(t_k,\nu_l)),
\label{eq:avscon}
\end{equation}
where $\Puv_{pqkl}$ is a boxcar-like window function that corresponds to bin $\Buv_{pqkl}$ in $uv$-space 
(and also includes the determinant term of eq.~\ref{eq2:conti:uv}). This makes it explicit that each averaged 
visibility is drawn from a convolution of the underlying visibilities with a boxcar-like window function.

Note what eq. (\ref{eq:avscon}) does and does not say. It does say that each individual averaged visibility corresponds to 
convolving the true visibilities by some window function. However, this window function is different for each baseline $pq$ and 
time/frequency sample $t_k,\nu_l$ (which is emphasised by the subscripts to $\Puv$ in the equations above). Averaging 
is thus not a ``true'' convolution, since the convolution kernel changes at every point in the $uv$-plane. \textcolor{black}{We will} call this 
process a \emph{pseudo-convolution}, and the kernel being convolved with ($\Puv_{pqkl}$) an example of a 
\emph{baseline-dependent window function} (BDWF). In subsequent sections we will explore alternative BDWFs.

In actual fact, a correlator (or an averaging operation in post-processing) deals with averages of discrete and noisy
samples, rather than a continuous integration. Ignoring the complexities of correlator implementation, let us cast
this process in terms of a simple averaging operation. That is, assume we have a set of \emph{hi-res} or
\emph{sampled visibilities} on a high-resolution time/frequency grid
$t_i,\nu_j$:
\begin{equation}
\label{eq:sampling}
\Vs_{pqij} = \VV_{pq}(t_i,\nu_j) + \NN[\sigma^\mathrm{(s)}_{pqij}],
\end{equation}
where $\VV_{pq}$ is given by eq.~(\ref{eq:visSky:2Dtf}), and $\NN$ represents the visibility noise term, which is
a complex scalar or complex $2\times2$ matrix with the real and imaginary parts being independently drawn from a 
zero-mean normal distribution with the indicated r.m.s. \citep{wrobel1999sensitivity}. The noise
term is not correlated across samples.
The \emph{lo-res} or \emph{averaged} or \emph{resampled} visibilities are then a discrete sum:
\begin{equation}
\label{eq:discrete:tf0}
\Vm_{pqkl} = \frac{1}{n} \sum_{ij\in\Bij_{kl}}  \Vs_{pqij},
\end{equation}
where $\Bij_{kl}$ is the set of sample indices $ij$ corresponding to the \emph{resampling bin}, i.e.
\begin{equation}
\Bij_{kl} = \big \{ ij:~t_i\nu_j \in \Btf_{kl} \big \},
\end{equation}
and $n = n_t\times n_\nu$ is the number of samples in the bin. 
Using the BDWF definitions above, this becomes a conventional discrete convolution (assuming a regular 
$t\nu$ grid):
\begin{equation}
\label{eq:discrete:tf}
\Vm_{pqkl} = \sum_{i,j=-\infty}^{\infty}  \Vs_{pqij} \Ptf(t_i-t_k,\nu_j-\nu_l).
\end{equation}
In $uv$-space, this becomes a discrete convolution on an irregular grid (the $\bmath{u}_{ij}$ grid being schematically illustrated by \ATM{Fig.~\ref{fig:uvcov}, left)}:
\begin{equation}
\label{eq:discrete:uv}
\Vm_{pqkl} = \sum_{i,j=-\infty}^{\infty}  \Vs_{pqij} \Puv_{pqkl}(\bmath{u}_{ij}-\bmath{u}_{kl}),
\end{equation}

\subsection{Effect of averaging on the image}
\label{sec:effectbw}
\begin{figure*}
\includegraphics[width=.4\textwidth]{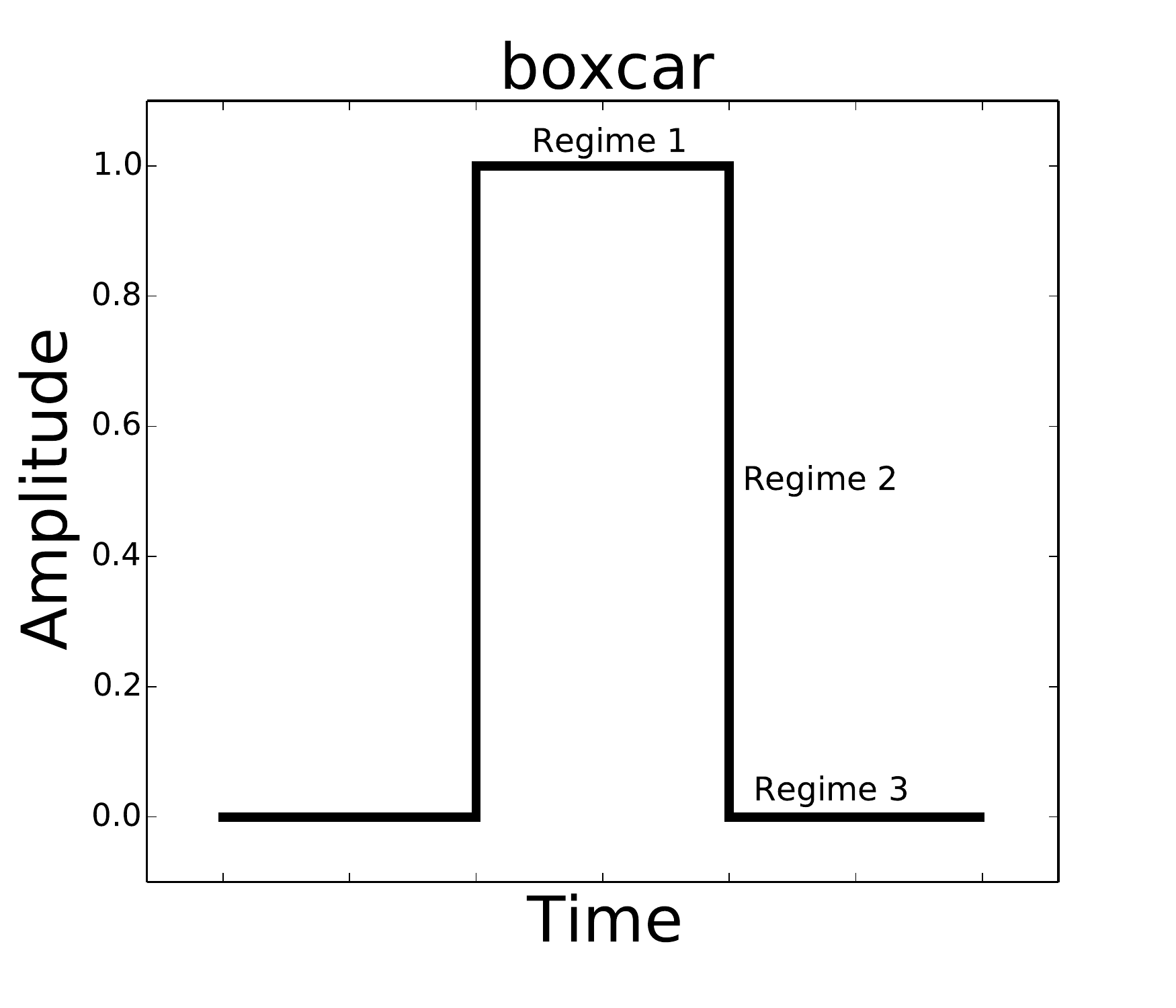}%
\includegraphics[width=.4\textwidth]{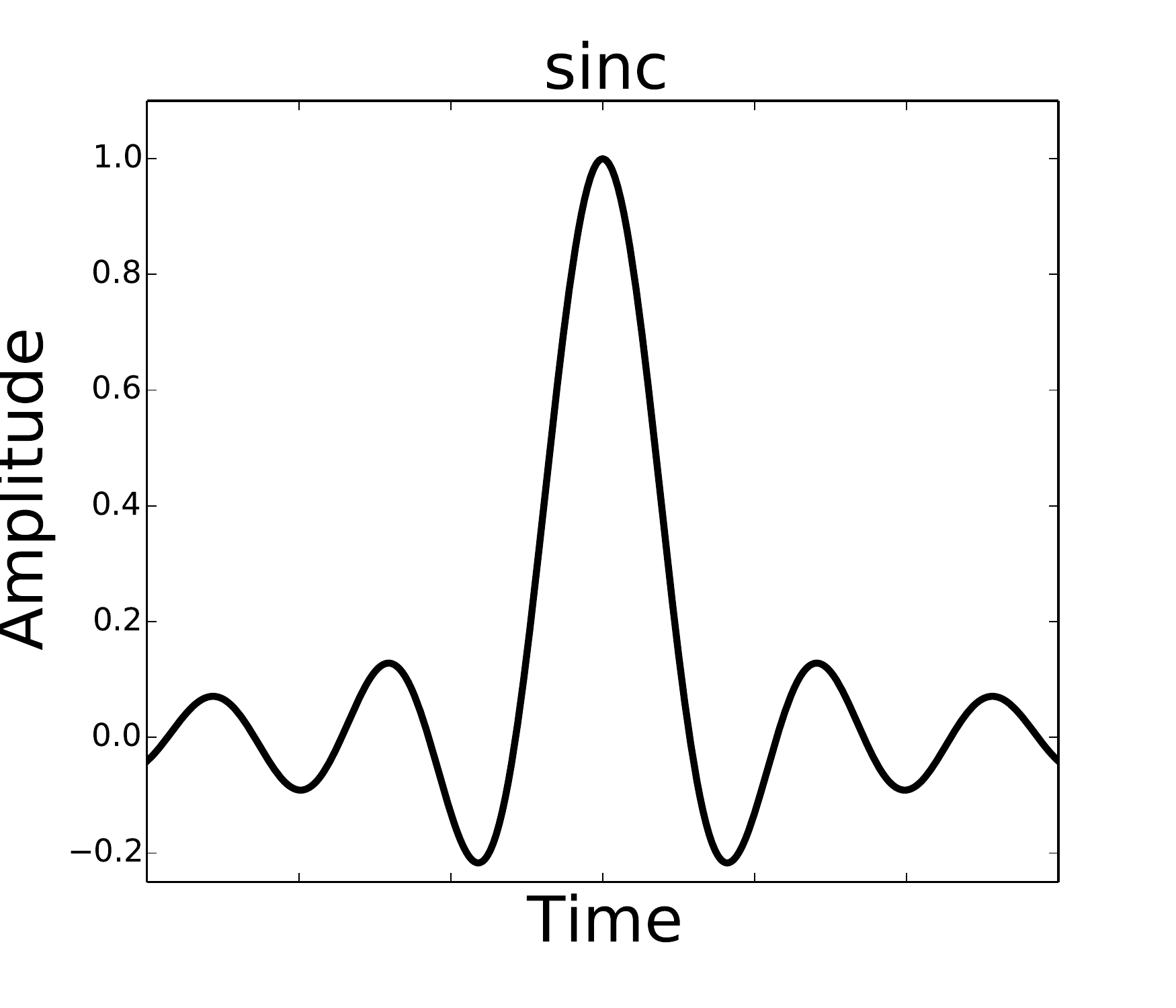}\\
\caption{Left: boxcar response. In the $uv$-plane, this represents the window function corresponding to normal
averaging of visibilities. In the image plane, this represents the ideal image-plane response function. Right: 
\Sincc~response. In the image plane, this represents the window response function corresponding to a boxcar window function in 
the $uv$-plane. In the $uv$-plane, this represents the ideal window function.}
\label{fig:idealwindowfunction}
\end{figure*}

In the limit of $\Delta t,\Delta \nu \rightarrow 0$, averaging becomes equivalent to sampling. 
An interferometer must, intrinsically, employ a finitely small averaging interval. The Fourier phase 
component $2\pi\phi(u,v,w)$ is a function of frequency and time, with increasing variation over the averaging interval 
for sources far from the phase centre. The average of a complex quantity with a varying phase then effectively ``washes out'' 
amplitude, the effect being especially severe for wide FoIs \citep[for an extensive discussion, see][]{bregman2012system}. In the
$uv$-plane, this
effect is often referred to as \emph{time} and \emph{bandwidth decorrelation}, and \emph{smearing} in the image plane.

The discussion above provides an alternative way to look at decorrelation/smearing. With averaging in effect, the relationship between the measured and the ideal visibility changes to (contrast this to eq.~\ref{eq:inst-sampled-visibility}): 
\begin{equation}
\VVM_{pqkl} = \delta_{pqkl} ( \VV\circ\Puv_{pqkl} ),
\end{equation} 
Combining this with eq.~\ref{eq:imaging3}, and using the Fourier convolution theorem, we can see that the dirty image is formed  as\textcolor{black}{:}
\begin{equation}
\IID =  \sum_{pqkl} W_{pqkl} \PP_{pqkl} \circ (\II\cdot\TT_{pqkl}),
\end{equation}
with the apparent sky $\II$ now tapered by the baseline-dependent \emph{window response function} $\TT_{pqkl}$, the latter being the inverse Fourier transform of the BDWF:
\begin{alignat}{2}
\TT_{pqkl} &= \FF^{-1}\{ \Puv_{pqkl} \}.
\end{alignat}
In other words, the dirty image yielded by averaged visibilities 
(compare this to the ideal dirty image given by eq.~\ref{eq:imaging:DI})
is a weighted average of per-visibility dirty images corresponding to a per-visibility tapered sky. The Fourier transform of a boxcar-like function is a \Sincc-like function, schematically illustrated in 1-D by Fig.~\ref{fig:idealwindowfunction} (right). Time and bandwidth smearing represents the average effect 
of all these individual tapers. Shorter baselines correspond to smaller boxcars and wider tapers, longer baselines to larger 
boxcars and narrower tapers, and are thus more prone to smearing.
\begin{figure*}
\includegraphics[width=\columnwidth]{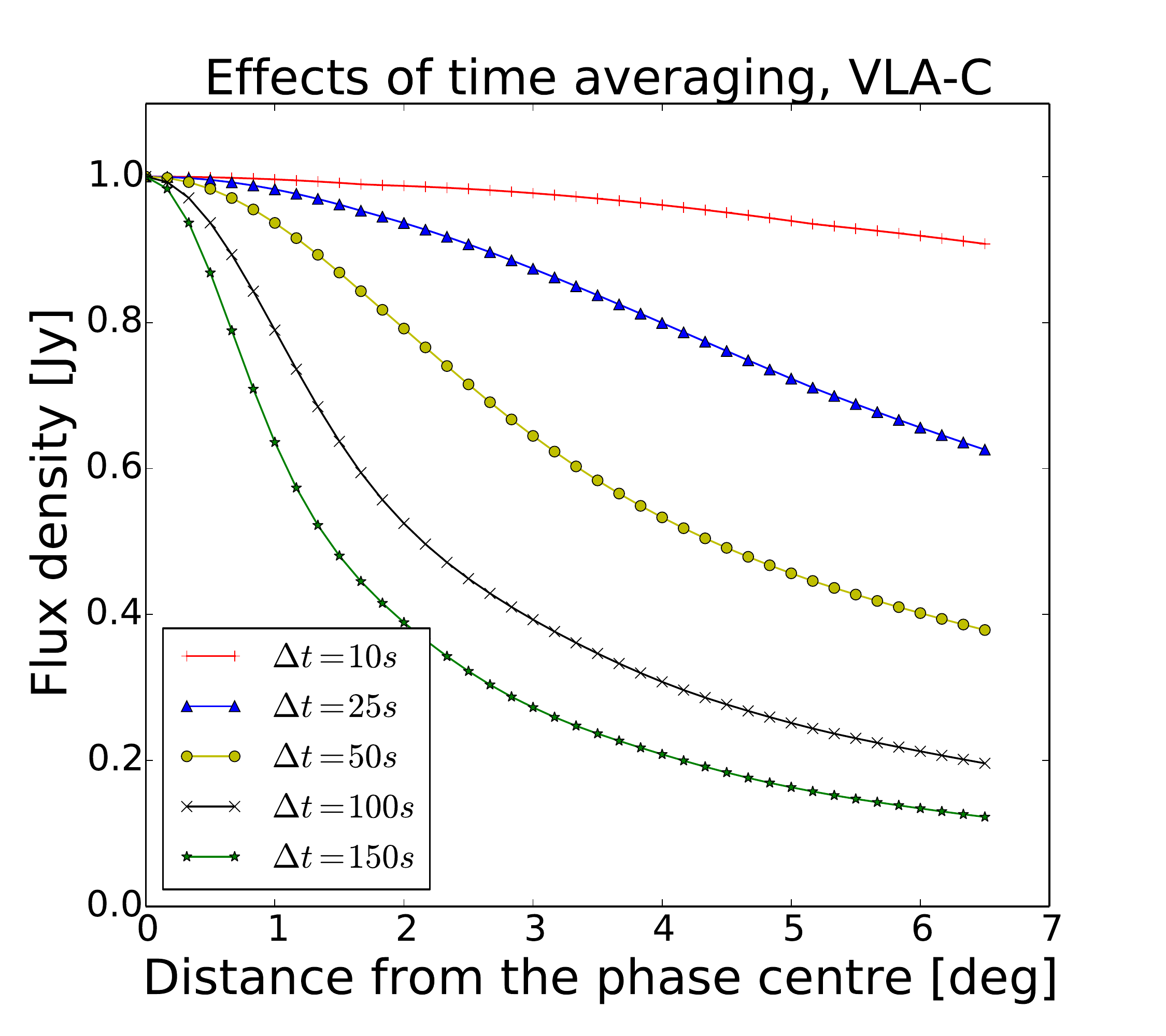}%
\includegraphics[width=\columnwidth]{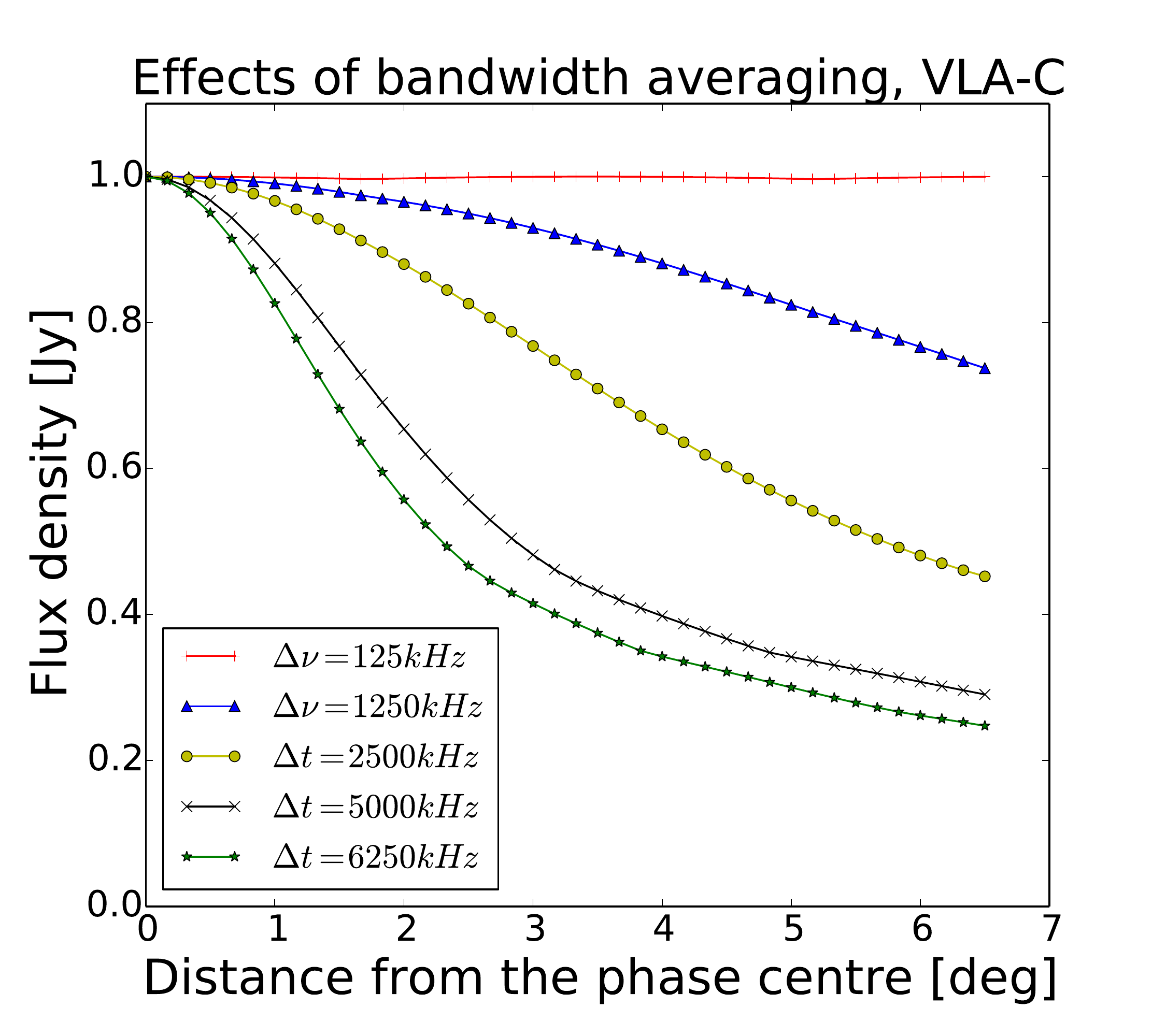}
\caption{Effects of time and frequency averaging: the apparent intensity of a 1 Jy source, as seen by \JVLA-C at 1.4 GHz, 
as a function of distance from phase centre. (Left) Frequency interval fixed at 125 kHz, time interval varies; 
(right) time interval fixed at 1 s, frequency interval varies.
}\label{fig:smear}
\end{figure*}
Fig.~\ref{fig:smear} (produced by simulating a series of high time-frequency resolution observation using MeqTrees~\citep{noordam2010meqtrees}, and 
applying averaging) shows the attenuation of a 1 Jy source as a function of distance
from phase centre, for a set of different time and frequency intervals. The simulations correspond to \JVLA~ 
in the C configuration, with an observing frequency of 1.4 GHz. At this frequency, the first null of the primary beam is at 
$r\approx36'$, and the half-power point is at $\sim16'$, thus we can consider the ``conventional'' FoI (i.e. the half-power 
beam width, or HPBW) to be about $0.5^\circ$ across. Note that the sensitivity of the upgraded \JVLA, as well as
improvements in calibration techniques \citep{Perley-3C147}, allow imaging to be done in the first primary beam sidelobe as well
(and in fact it may be necessary for deep pointing, if only to deconvolve and subtract sidelobe sources), so we could also
consider an ``extended'' FoI out to the second null of the primary beam at $r\approx1.25^\circ$. Whatever definition
of the FoI we adopt, Fig.~\ref{fig:smear} shows that to keep amplitude losses across
the FoI to within some acceptable threshold, say 1\%, the averaging interval cannot exceed some critical size,
say 10 s and 1 MHz. Conversely, if we were to adopt an aggressive averaging strategy for the purposes of data 
compression, say 50 s and 5 MHz, the curves indicate that we would suffer substantial amplitude loss towards the 
edge of the FoI. 

Finally, note that the curves corresponding to acceptably low values of smearing across the FoI (i.e. up to 25 s and 
up to 1.25 MHz) have a very gentle slope, with very little suppression of sources \emph{outside} the FoI. 

\subsection{The case for alternative BDWFs}
The window response or image plane response (IPR) function induced by normal averaging (Fig.~\ref{fig:idealwindowfunction} (right)) is far from ideal: it either suppresses too much within 
the FoI, or provides limited benefit to suppressing outside the FoI sources, or both. The optimal image plane response would be 
a  disk-like function, unity within the region  bounded by a circle and zero outside this region. In 1-D, it is a 
boxcar-like, as in Fig.~\ref{fig:idealwindowfunction} (left).

The BDWF that would produce such a response is \Sincc-like, as in 1-D presented in Fig.~\ref{fig:idealwindowfunction} (right). 
The problem with a \Sincc~is that it has infinite support; applying it over finite-sized bins necessarily means a \emph{truncated} 
BDWF that results in a sub-optimal taper. The problem of optimal filtering has been well studied in signal processing (usually
assuming a true convolution rather than the pseudo-convolution we deal with here), and we shall apply these lessons below.

The derivations above make it clear that using a different BDWF in place of the conventional boxcar-like $\Puv$ 
could in principle yield a more optimal tapering response. The obvious \textcolor{black}{disadvantage} is a loss in 
sensitivity. Each visibility sample is subject to an independent Gaussian noise term in the real and imaginary part; the noise of
the average of a set of samples is minimised when the average is naturally weighted (or unweighted, if the noise is 
constant across visibilities). Thus, any deviation from a boxcar F must necessarily increase the noise in the 
visibilities. Below we will study this effect both theoretically and via simulations, to establish whether this 
trade-off is sensible, and under which conditions.

\section{Applying window functions to visibilities}
While visibilities are (usually) regularly sampled in $t\nu$-space, in $uv^\mathrm{m}$-space this is not so. In frequency, 
the sampling positions go as $\sim\lambda^{-1}$, while in time, baselines with a longer East-West component sweep out longer tracks between successive 
integrations (Fig.~\ref{fig:uvcov}). Applying a window function with a constant integration window in $t\nu$ space corresponds to 
different-sized windows in $uv$-space. In the case of normal averaging, this results in the boxcar-like window $\Puv$ of 
eq.~(\ref{eq:avscon}) having a baseline-dependent scale. The scale of the tapering response being inversely proportional to 
the scale of the window function, this results in more decorrelation (i.e. a narrower achievable FoI) on longer baselines.

\begin{figure*}%
\centering
\includegraphics[width=.4\textwidth]{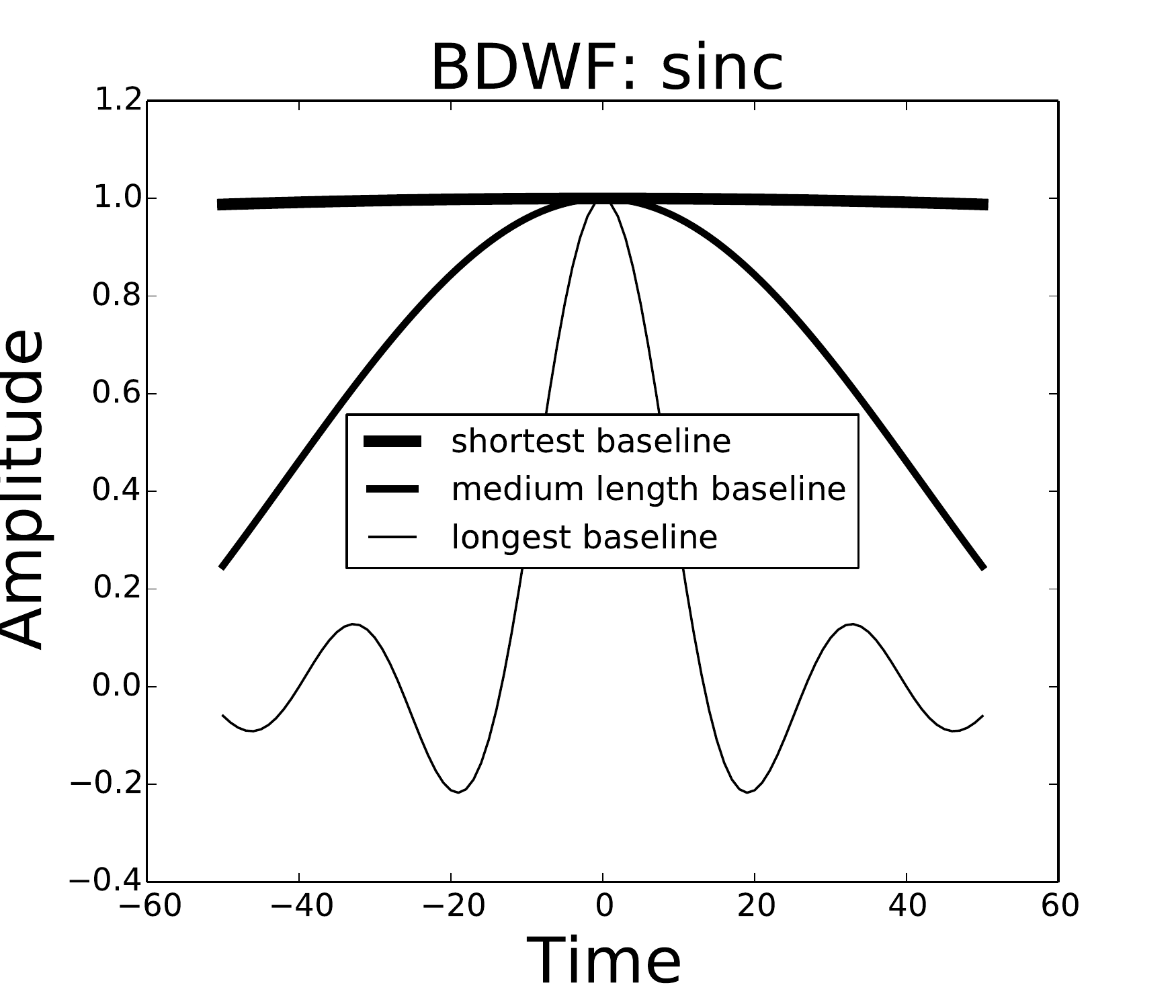}%
\includegraphics[width=.4\textwidth]{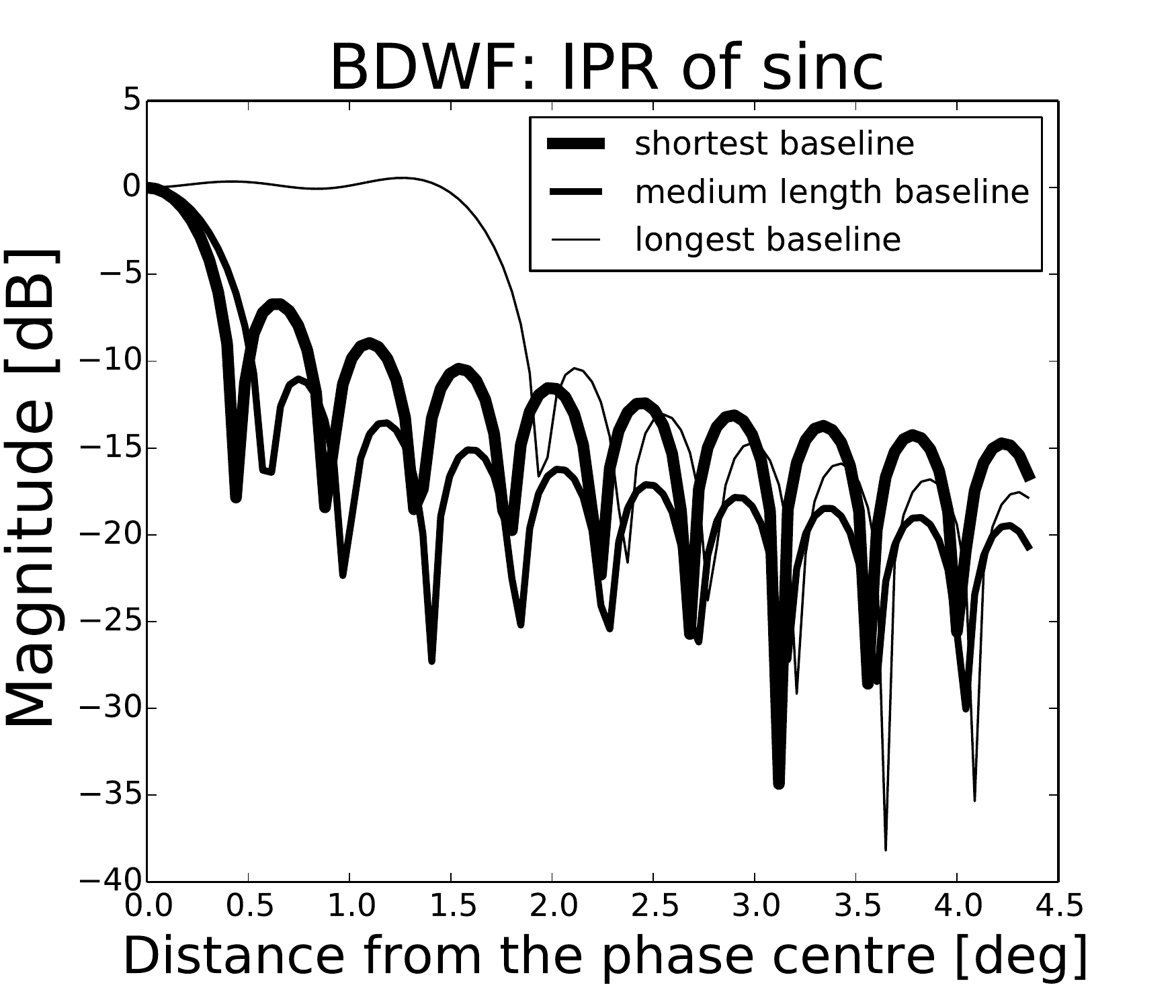}\\
\includegraphics[width=.4\textwidth]{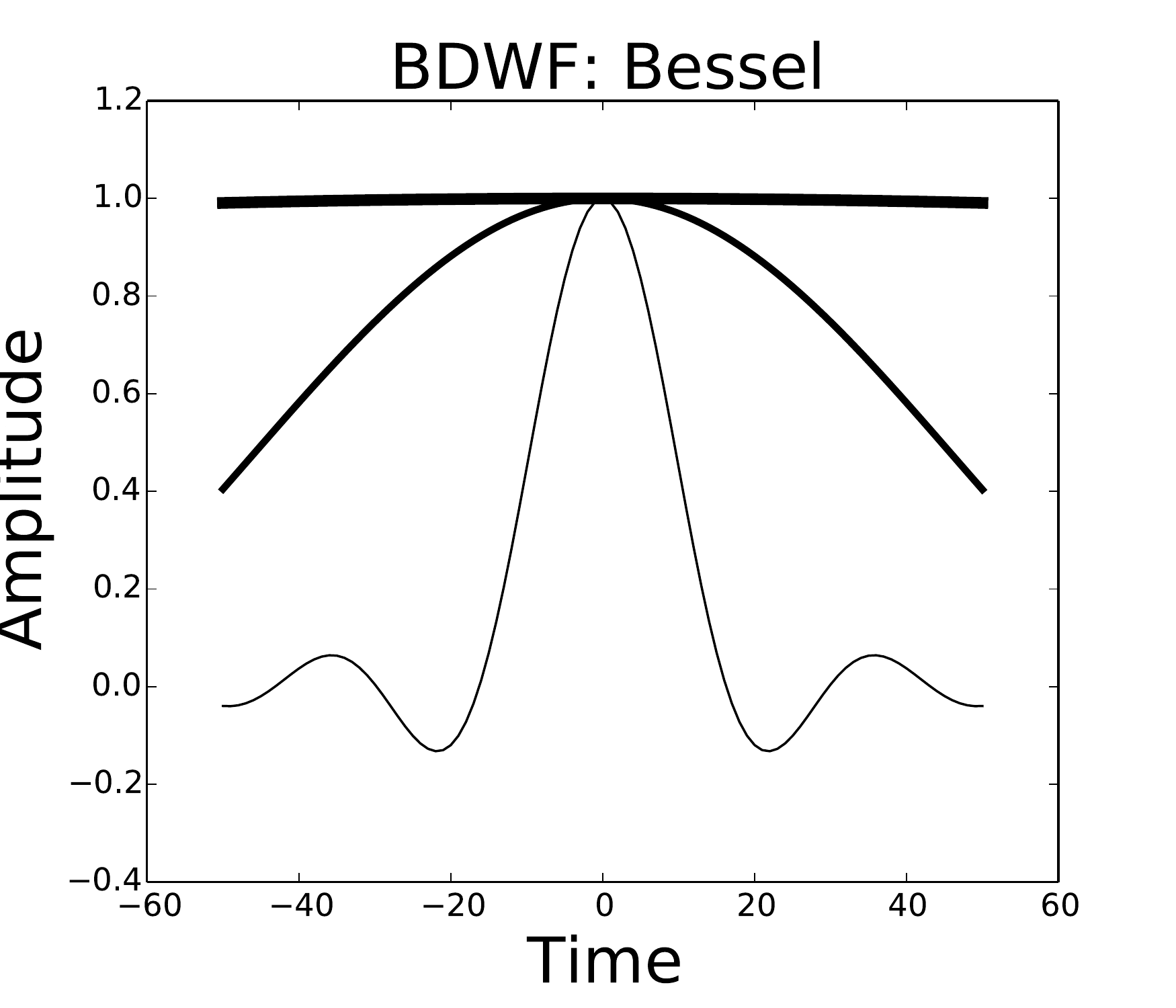}%
\includegraphics[width=.4\textwidth]{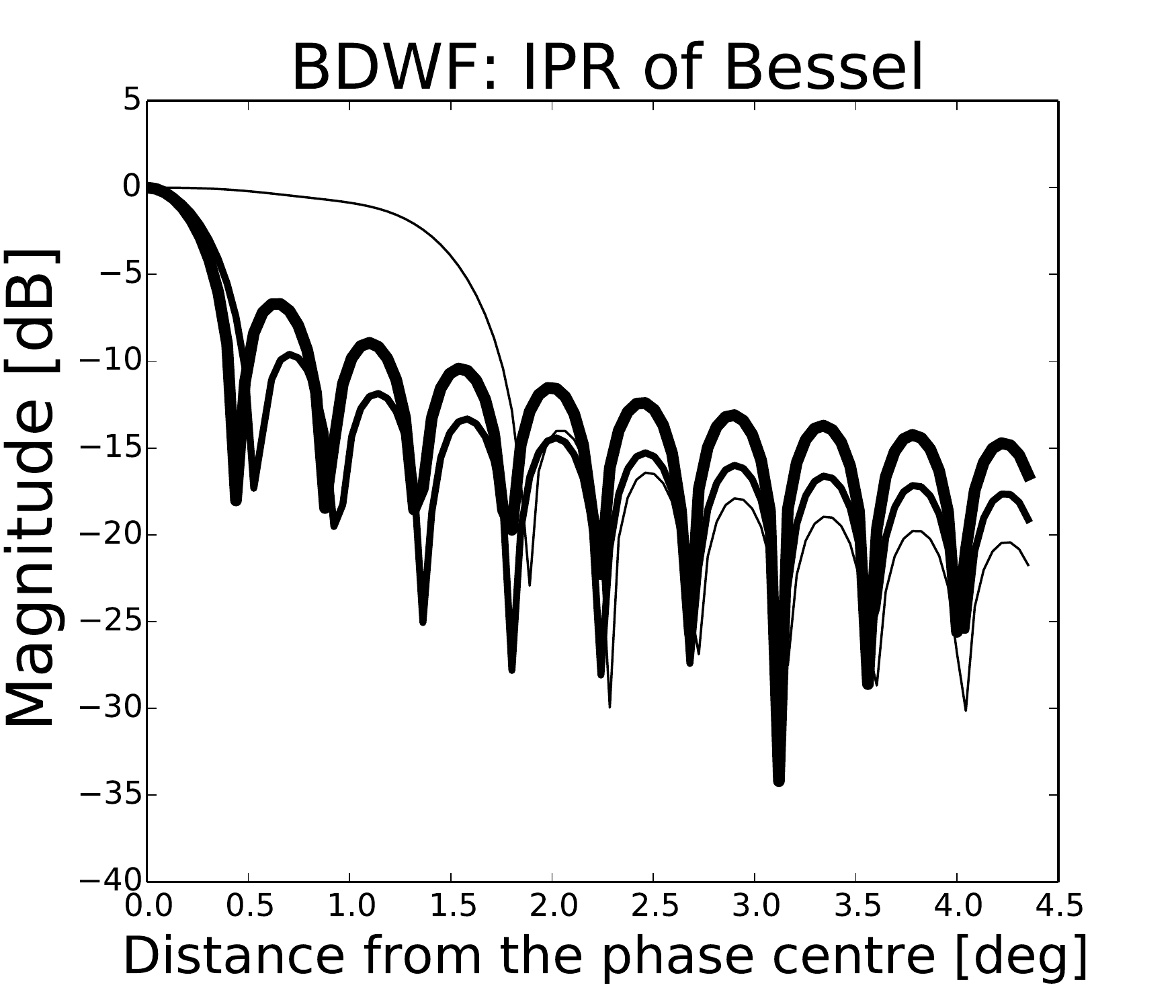}
\caption{Cross-sections through two different BDWFs (top left: $\mathrm{sinc}$, bottom left: Bessel) defined in $uv$-space, 
plotted along the time axis and their IPR (top right: IPR for $\mathrm{sinc}$, bottom right: IPR for Bessel). This shows that the effective window function is a scaling and truncation of the underlying window function, with
the shortest baselines reducing to a boxcar-like window function.}
\label{fig:WF:perbaseline}
\end{figure*}

By defining our alternative window functions in $uv$-space (in units of wavelength), we can attempt to ``even out'' the decorrelation 
response across baselines. For a given BDWF $X(u,v)$, we have the following recipe 
for computing resampled visibilities (compare to eq.~\ref{eq:discrete:uv}):
\begin{equation}
\label{eq:avg:wf}
\Vm_{pqkl}= \frac{\displaystyle\sum\limits_{{i,j}\in \Bij_{kl}} \Vs_{pqij} X(\bmath{u}_{pqij}-\bmath{u}_{pqkl})}
{\displaystyle\sum\limits_{{i,j}\in \Bij_{kl}} X(\bmath{u}_{pqij}-\bmath{u}_{pqkl})},
\end{equation}
where $\bmath{u}_{pqkl}$ is the midpoint of the resampling bin $\Bij_{kl}$ in $uv$-space. The main lobe of the window function then 
has the same scale across the entire $uv$-plane, while the resampling bins have different $uv$-sizes. Conversely, in 
$t\nu$-space the bins
are regular, while the main lobe of the effective window function scales inversely with the baseline fringe rate. Furthermore, 
the window function is truncated at the edge of each bin; on the shortest baselines this truncation is extreme to the 
point of approaching the boxcar-like $\Puv$ (Fig.~\ref{fig:WF:perbaseline}).
% \begin{figure*}%
% \includegraphics[width=.4\textwidth]{./Figures/sinc-baseline_longMidShortshortgrey.pdf}%
% \includegraphics[width=.4\textwidth]{./Figures/bessel-baseline_longMidShortshortgrey.pdf}
% \caption{Cross-sections through two different BDWFs (left: \Sincc, right: Bessel) defined in $uv$-space, 
% plotted along the time axis. This shows that the effective window function is a scaling and truncation of the underlying window function, with
% the shortest baselines reducing to a boxcar-like window function.
% }
% \label{fig:WF:perbaseline}
% \end{figure*}
The downside of this simple approach is twofold. Firstly, while all  the window functions above nominally exhibit far lower 
sidelobes than the boxcar
(i.e. more suppression for out-of-FoI sources), they no longer perform \textcolor{black}{particularly} well under truncation, 
with extremely truncated window functions 
at the shorter baselines becoming boxcar-like. Secondly, taking a weighted sum in eq.~\ref{eq:avg:wf} increases 
the noise in comparison to normal averaging.

% \textcolor{red}{ Reviewer comments Sections:
%      1.1) Section 3 was confusing. The title suggests making a connection
%           to DSP theory which is not that important to have it's own section. It
%           then turns out that this is where you generalize the standard one
%           dimensional tapers to 2D tapers, am I correct?
%           Suggest removing DSP parts and moving the 2D taper generalizations to the
%           definitions of $X_pqkl$, eq 36, in terms of \Sincc and Bessel functions.
%           (You really need explicit definitions of the BDWF you use.) Also try
%           to motivate why you chose these and not other functions.
%      1.2) The appendix, which is 3 pages long, does not provide more than
%           a basic summary of windowing/taper functions. Suggest removing it completely.
%           (Especially as you do not even use most of these functions in the main text.)
%           It suffices to have a reference to a signal processing text, or may be save just
%           the Table.}
\ATM{We use the standard discrete signal processing (DSP) filter terminology to describe the IPR of visibility domain
window functions. Window functions  or rather their corresponding image-plane response (IPR) can be characterised in terms of various metrics. Some common  ones are the peak sidelobe level, the main lobe width and the sidelobes roll-off  rate \citep{smith1997scientist}. In terms of the ``ideal'' IPR (Fig.~\ref{fig:idealwindowfunction}, left), these correspond to the following desirable traits:
\begin{itemize}
\item Maximally conserve the signal within the FoI (``\textcolor{black}{Regime 1}'' in the figure),
and make the transition region (``\textcolor{black}{Regime 2}'') as sharp as possible. Both of these correspond to larger main lobe width.
\item Attenuate sources outside the FoI (``\textcolor{black}{Regime 3}''): this corresponds to a lower peak sidelobe level and higher sidelobes roll-off.
\end{itemize}
Table~\ref{tab:WF:performance} (see Appendix~\ref{appendixA}) summarises the performance of  different window functions well known in DSP~\citep[]{smith1997scientist,prabhu2013window}.  
This table shows that the IPR of the $\mathrm{sinc}$, the Bessel of the first kind of order zero ($J_0$)~\citep{watson1995treatise} and all their derivatives with Hamming, Han and Blackman window functions have a larger main lobe,  low peak sidelobe level and high sidelobe roll-off compared to others. Hence, they provide the most promising IPRs for our purposes. Note that any window derived from the $\mathrm{sinc}$ or Bessel with Hamming, Han or Blackman requires two successive visibility weightings, which may results in amplifying the thermal noise  compared to the $\mathrm{sinc}$ or Bessel. This makes the $\mathrm{sinc}$ and the Bessel more suitable for this work.
For this reason, we have chosen to use the $\mathrm{sinc}$ and the Bessel  window functions to serve as the basis of BDWFs
developed in the rest of this paper. 
We use the following definitions to construct a 2-D  $\mathrm{sinc}$ and Bessel window functions from their 1-D variants: 
%To construct two-dimensional  window functions from one-dimensional window functions, we will use the following definitions:
\begin{alignat}{2}
X(u,v) &= X(\kappa u)X(\kappa v), ~~ X\equiv \mathrm{sinc}  \\
X(u,v) &= X(\kappa r),~r=\sqrt{u^2+v^2}, ~~ X\equiv J_0.
\end{alignat}
Here the FoI
is adjustable by the parameter $\kappa$ (in radians).}
\subsection{Overlapping BDWFs}
\label{baseline2}
\newcommand{\WF}[3]{{#1}-$#2${}$\times${}$#3$}
\begin{figure} 
\includegraphics[width=\columnwidth]{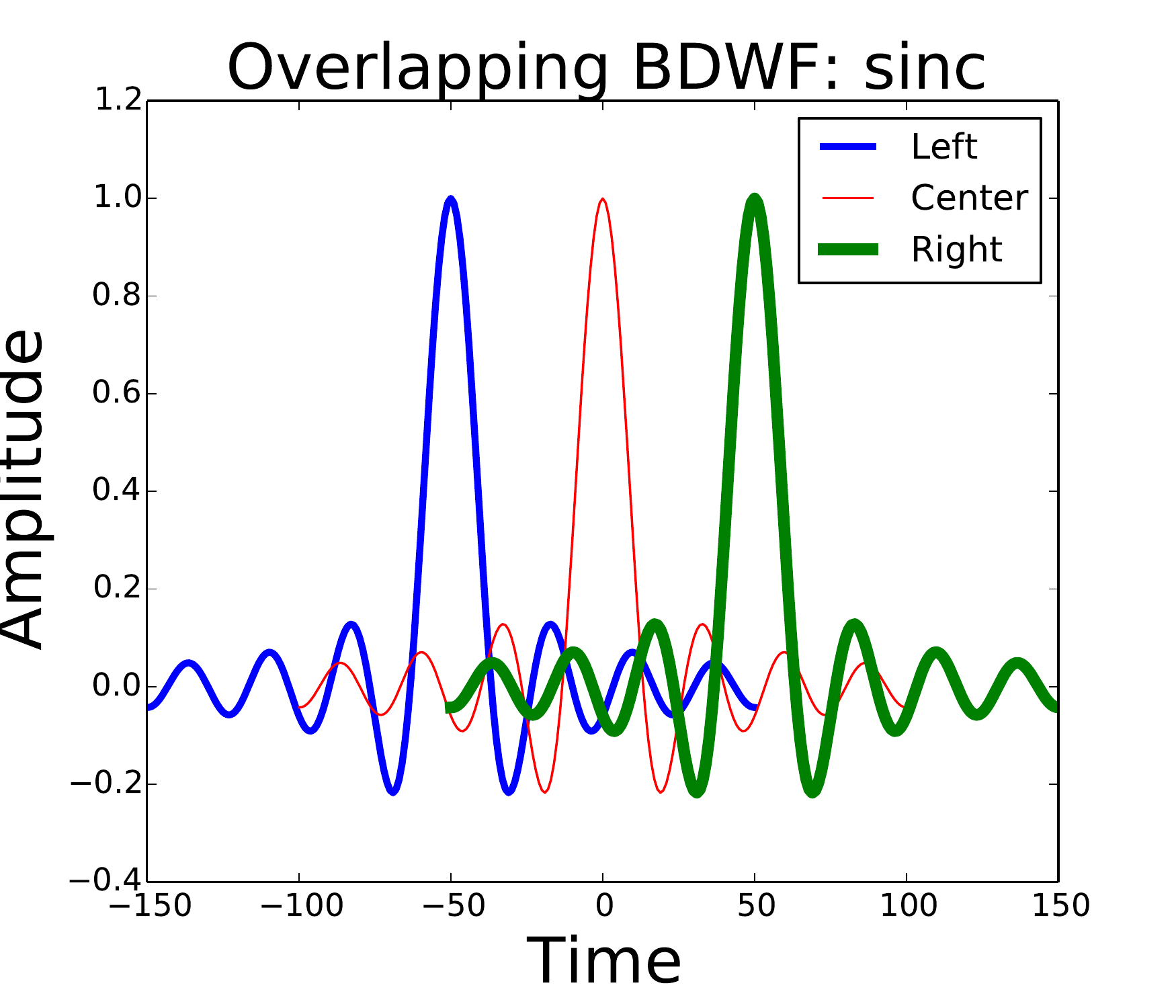}
\caption{Overlapping 
BDWFs representing adjacent resampling bins (\ATM{right resampling bin, centre resampling bin and left resampling bin}). This corresponds to overlap factor $\alpha=3$ in time.
}\label{fig:overlap}
\end{figure}
A more sophisticated approach involves overlapping BDWFs. Normal averaging implicitly assumes that the resampling bins $B_{kl}$ 
in eq.~(\ref{eq:avg:wf}) \textcolor{black}{do not} overlap for adjacent $kl$, since they represent adjacent averaging intervals.
There is, however, no reason (apart from computational load) why we cannot take the sum in eq.~(\ref{eq:avg:wf}) 
over larger bins. Let us define the \emph{window bin} for \emph{overlap factors} of $\alpha,\beta$ as
\begin{equation}
\Bab_{kl} = \big \{ ij:~t_i\nu_j \in \Babtf_{kl} \big \},\label{eq:windowbinalphabeta}
\end{equation}
i.e. as the set of sample indices corresponding to a bin of size $\alpha\Delta t\times\beta\Delta \nu$ in $t\nu$-space.
Let us then compute the sums in eq.~(\ref{eq:avg:wf}) over the window bin. This becomes distinct from the
resampling bin: while the latter represents the spacing of the resampled visibilities, the former represents the size 
of the window over which the convolution is computed. Only for $\alpha=\beta=1$ do the two bins become the same.

% \textcolor{red}{Reviewer comments: 5) You mention the computational complexity in passing. I think it would
%     be useful to have rough estimate of the computational cost, seeing as you
%     make an explicit mention of the compression factor.}   
\ATM{ 
As  mentioned above, the  computational load for overlapping window functions will increase linearly as the factors $\alpha$ and $\beta$ increase. For a given  resampling bin $\Bab_{kl}$ the computational 
complexity would scale as
$\mathcal{O}(\alpha \beta n_{t}n_{\nu})$ compared to $\mathcal{O}(n_{t}n_{\nu})$, which is the complexity for applying a non-overlapping
window function to visibilities within the resampling bin $\Btf$. In this notation, the
resampling bin for a non-overlapping window functions consists of $n_{t}\times n_{\nu}$ samples, which is the number of 
multiplications between the baseline-dependent weights and the visibilities.}

\ATM{In the overlapping regime, the baseline-dependent weight for a single  visibility is not 
defined by a unique BDWF,
but by the strength of the correlation between the overall
overlapping BDWFs with the visibility. 
BDWFs in the overlapping regime are schematically illustrated in Fig.~\ref{fig:overlap}.
% \ATM{The weight of a uv-bin is not defined by a unique BDWF
% but by the strength of the correlation between the overall
% overlapping BDWF as shown in Figure ??. That said, the
% weight of a uv-bin becomes the summation of overlapping
% BDWF samples on the uv-bin normalized by the summation
% of the overall uv-bins weights within the range where the
% overlap samples are defined.
For simple averaging, 
overlapping offers no benefit, since it only broadens $\Puv$ and therefore increases smearing, but for a well-behaved 
BDWF, enlarging the window bin (while maintaining the same window function scale) means less truncation -- thus lower 
sidelobes -- and decreased noise, as more sampled visibilities are taken into account. On longer baselines, 
the IPR of a well-behaved overlapping BDWF  means less smearing in the FoI and 
excellent out-of-FoI suppression (see Fig.~\ref{fig:overlap-regime}). 
%This behaviour is depicted in Figure~\ref{}.
On  shorter baselines, a well-behaved BDWF is equivalent to a boxcar and therefore enlarging the window size by overlapping results in a
 decrease  of noise (see Sect.~\ref{sec:imaging}).}

\ATM{Appendix~\ref{appendixB} provides an alternative way to look at the overlapping BDWFs applied to visibilities.}
% Overlapping BDWFs are schematically illustrated in Fig.~\ref{fig:overlap}. For normal averaging 
% overlapping offers no benefit, since it only broadens $\Puv$ and therefore increases smearing. But for a well-behaved 
% BDWF, enlarging the window bin (while maintaining the same window function scale) means less truncation -- thus lower 
% sidelobes -- and decreased noise, as more sampled visibilities are taken into account.

To distinguish overlapping BDWFs from non-overlapping ones, in the rest of the paper we will designate the 
window functions employed as \WF{\em WF}{\alpha}{\beta}. For example, \WF{\Sincc}{3}{2}, \WF{$J_0$}{1}{1} (i.e. no overlap), etc.
If resampling is only done in one direction (only time or only frequency), \textcolor{black}{we will} indicate this as e.g. \WF{$J_0$}{3}{-}.

\subsection{Noise penalty estimates: analytic}
\label{sec:imaging}
Let us now work out analytically the \emph{noise penalty} associated with replacing an unweighted average by 
a weighted sum. For simplicity, \textcolor{black}{let us} assume that the noise term
has constant r.m.s. $\sigma_\mathrm{s}$ across all baselines and samples. If the resampling bin 
consists of $n_\mathrm{avg} = n_t\times n_\nu$ samples, and since the noise is not correlated between samples, 
the noise on the averaged visibilities in eq.~(\ref{eq:discrete:tf0}) will be given by
\begin{equation}
\sigma_\mathrm{m}^2 = \frac{1}{n_\mathrm{avg}^2} \sum_{i=1}^{n_\mathrm{avg}} \sigma_\mathrm{s}^2  = \frac{\sigma_\mathrm{s}^2}{n_\mathrm{avg}}
\end{equation}
Note that the noise is uncorrelated across averaged visibilities. We can therefore use the
the imaging equation (\ref{eq:imaging3}) to derive the following expression for the variance of the noise 
term in each pixel of the dirty image:
\begin{equation}
\label{eq:noise:image}
\sigma_\mathrm{pix}^2 = \frac{ (\sum_{pqkl} W_{pqkl}^2 \sigma_\mathrm{m}^2) }{ (\sum_{pqkl} W_{pqkl})^2 },
\end{equation}
which for natural image weighting ($W\sim\sigma^{-1}_\mathrm{m}$, i.e. $W\equiv1$ in this case) is simply
\begin{equation}
\sigma_\mathrm{pix}^2 = \frac{1}{N}\frac{\sigma_\mathrm{s}^2}{n_\mathrm{avg}},
\end{equation}
where $N$ is the total number of visibilities used for the synthesis.

To simplify further notation, \textcolor{black}{let us} replace $pqkl$ by a single index $\mu$, enumerating all the lo-res visibilities $\Vm_\mu$, with 
$\mu=1\dots N$. If we now employ eq.~(\ref{eq:avg:wf}) to compute the lo-res visibilities using some BDWF $X(u,v)$, the noise 
term becomes different per each visibility $\mu$:
\begin{equation}
\label{eq:noise:bdwf}
\sigma_{X\mu}^2 = \frac{\sum X^2(\bmath{u}_{pqij}-\bmath{u}_{pqkl})}
{\big [ \sum X(\bmath{u}_{pqij}-\bmath{u}_{pqkl}) \big ]^2 } \, \sigma_\mathrm{s}^2
\end{equation}
where both sums are taken over the window bin, ${i,j}\in \Bij_{kl}$. Let us define 
the \emph{visibility noise penalty} associated with BDWF $X$ 
and visibility $\mu$ as the relative increase in noise over the unweighted average, i.e.
\begin{equation}
\Xi_{X\mu} = \frac{\sigma_{X\mu}}{\sigma_\mathrm{m}} = 
\frac{ \sqrt{ n_\mathrm{avg} \sum X^2(\bmath{u}_{pqij}-\bmath{u}_{pqkl}) }}
{\sum X(\bmath{u}_{pqij}-\bmath{u}_{pqkl}) }. 
\end{equation}
Note that in the case of overlapping BDWFs, the window bin in eq.~\ref{eq:noise:bdwf} is larger than the 
resampling bin, and contains $n_X$ samples, with $n_X=\alpha\beta n_\mathrm{avg}$, where $\alpha$ and $\beta$ are the 
overlap factors. For $\alpha=\beta=1,$ it is easy to see that $\Xi_{X\mu}\geq1$, and only reaches 1 when $X\equiv1$. 
In other words, non-overlapping BDWFs always result in a visibility noise penalty above 1, while overlapping BDWFs 
can actually \emph{reduce} noise in the resampled visibilities. 

While paradoxical at first glance, this reduction in noise does {\bf not} result in a net gain in image 
sensitivity. The reason for this is that with overlap in effect, the noise terms become correlated across 
resampled visibilities $kl$ (within the same baseline $pq$), with each hi-res visibility sample contributing to 
multiple resampled visibilities, and the image noise term no longer follows eq.~(\ref{eq:noise:image}).

If the resampled visibilities correspond to a single-channel snapshot, or if the BDWFs are non-overlapping,
then the noise across visibilities remains uncorrelated, and we can compute the \emph{image noise penalty} 
associated with imaging weights $W$ and BDWF $X$ as
\begin{equation}
\Xi^W_X = 
\frac{\sigma_{\mathrm{pix},X}^2}{\sigma_{\mathrm{pix}}} = \frac{n}{\sigma_\mathrm{s}}
\frac{ (\sum_{\mu} W_{\mu}^2 \sigma_\mu^2) }{  (\sum_{\mu} W_{\mu})^2 } =
\frac{ (\sum_{\mu} W_{\mu}^2 \Xi_{X\mu}^2) }{ (\sum_{\mu} W_{\mu})^2 } 
\end{equation}
In the case of natural weighting ($W_\mu=\sigma_\mu^{-1}$) this reduces to:
\begin{equation}
\label{eq:noisepenalty:natural}
\Xi_X^{\mathrm{nat}} = \frac{N}{ \sum_{\mu} \Xi_{X\mu}^{-1} }.
\end{equation}

\subsection{Noise penalty estimates: empirical}
\label{subsec:noise}
In this section we employ simulations to empirically verify noise estimates computed using the derivation 
above. We generate a high resolution (``high-res'')
\JVLA-C measurement set (MS) corresponding to a 400 s synthesis with 1 s integration, with 30 MHz of bandwidth
centred on 1.4 GHz, divided into 360 channels of 83.4 kHz each. The MS is filled with simulated thermal noise 
with $\sigma_\mathrm{s}=1$ Jy. We then generate a low resolution (``low-resolution'') MS using 100 s integration, with 
a single frequency channel of 10 MHz. This MS receives the resampled visibilities. 
The size of the resampling bin is thus 100 s by 10 MHz, or $100\times120$ in terms of the number of hi-res samples.

We then resample the hi-res visibilities using a number of BDWFs, and store the results in the lo-res MS:
\begin{itemize}
\item Standard averaging to 100 s and 10 MHz (using the middle 120 channels). This gives us the baseline noise
estimate.
\item \Sincc~ and Bessel windows using the same bin, without an overlap, tuned to a FoI of $2^\circ$.
\item The same windows with overlap factors of $4\times3$.
\end{itemize}
We then image the lo-res MS and take the r.m.s. pixel noise across the image as an estimator of $\sigma_\mathrm{pix}$,
divide it by the baseline estimate produced with normal averaging, and compare the resulting noise penalty with that
predicted by eq.~\ref{eq:noisepenalty:natural}. Note that this estimator is not perfect, since image noise is correlated
across pixels. Nonetheless, we obtain results that are broadly consistent with analytical predictions 
(Table~\ref{tab:noise-comparison}).

\begin{figure*}
\includegraphics[width=.4\textwidth]{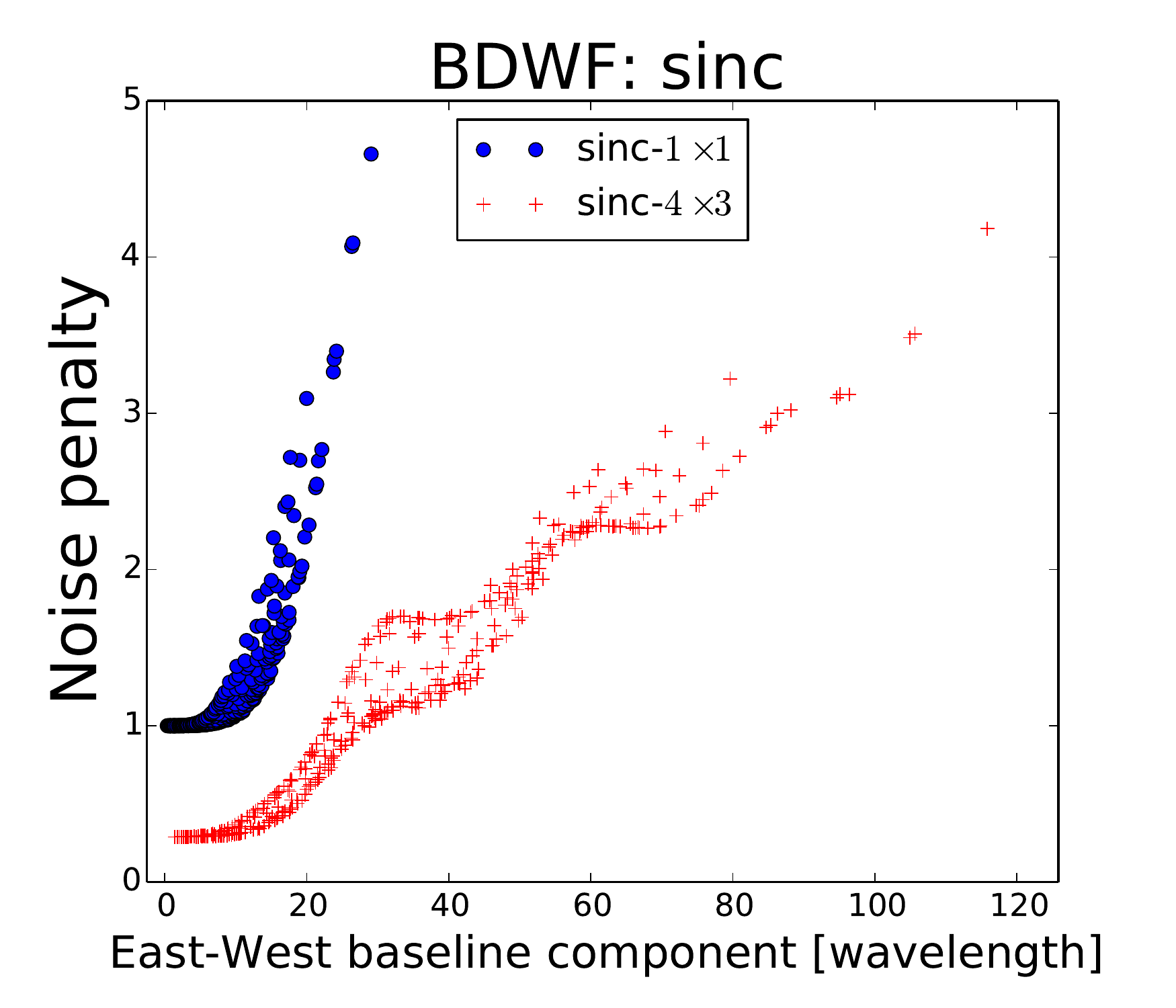}%
\includegraphics[width=.4\textwidth]{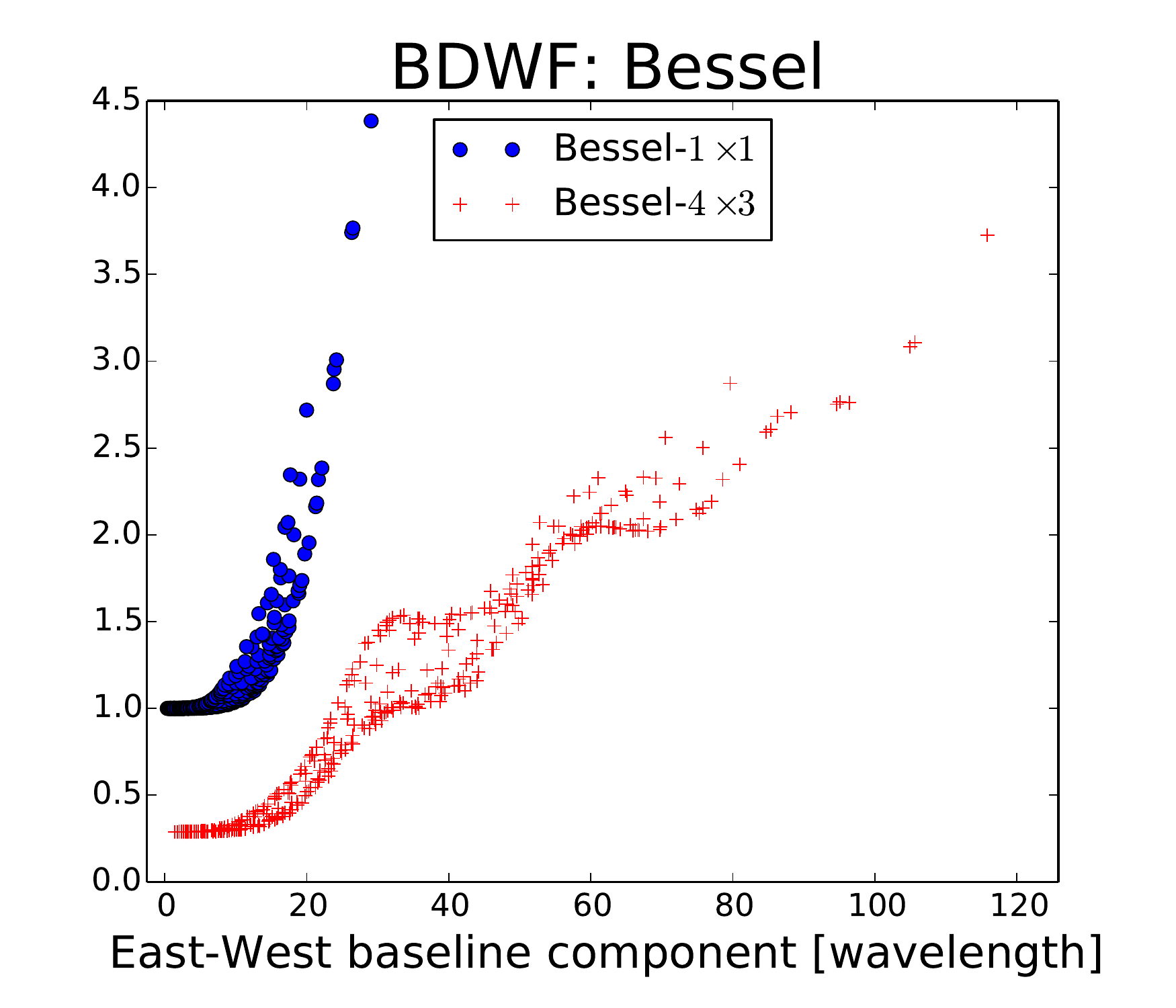}
\caption{Noise penalty w.r.t. normal averaging as a function of baseline length for two different BDWFs 
(\Sincc~and Bessel), with and without overlap. \JVLA-C, 1.4 GHz, sampling intervals of 100 s and 
10 MHz. 
}
\label{fig:per-baseline-noise}
\end{figure*}

Fig.~\ref{fig:per-baseline-noise} shows the predicted visibility noise penalty for the same BDWFs, as a function of 
East-West baseline component, which determines the baseline rotation speed. Note that the noise penalty rises sharply towards 
longer Eeat-West baselines. Note also that the penalty is well below 1 on shorter baselines, when overlapping is in effect.
\ATMNEW{As mentioned in Sect.~\ref{baseline2}, this is because the BDWF 
becomes equivalent to boxcar averaging (i.e. unweighted averaging or simple averaging) and can only decrease the noise when overlap
is in place. This is simple to understand analytically as follows: the noise term has constant r.m.s. $\sigma_\mathrm{s}$ across all baselines (including samples) and  the resampling bin consists of $n_t\times n_{\nu}$ samples. Since the noise is not correlated between samples, the noise
on the averaged (unweighted averaging) visibilities in  eq.~(\ref{eq:discrete:tf0}) is given by:
\begin{alignat}{2}
\sigma_\mathrm{m}^2&=\frac{\sigma_\mathrm{s}^2}{n_tn_{\nu}}.
\end{alignat}
On shorter baselines, the BDWF $X\equiv 1$ (i.e. unweighted averaging) and overlapping means 
enlarging the  averaged resampling bin by a factor of $\alpha\beta$ with $\alpha > 1$ and/or $\beta > 1$. The noise on the averaged visibilities becomes:
\begin{equation}
\sigma_{X\mu}^2 = \frac{\sigma_\mathrm{s}^2}{\alpha \beta n_tn_{\nu}},
\end{equation}
which gives us a noise penalty of 
\begin{equation}
\Xi_{X\mu} = \frac{\sigma_{X\mu}}{\sigma_\mathrm{m}} = \frac{1}{\sqrt{\alpha \beta}} < 1.
\end{equation}
Note that as $\alpha\beta >1$, it implies that $\sqrt{\alpha\beta} >1$.}
\begin{table}
\begin{tabular}{lll}
\hline
{\bf BDWF} & {\bf $\Xi$ analytic } & {\bf $\Xi$ sim}\\
\hline\hline
\WF{\Sincc}{1}{1} &$1.247$ &$1.22$\\
\WF{\Sincc}{4}{3} &$1.242$ &$1.20$\\
\hline
\WF{\Bessel}{1}{1} & $1.178$& $1.16$\\
\WF{\Bessel}{4}{3} & $1.109$& $1.12$\\
\hline
\end{tabular}
\caption{A comparison of image noise penalties associated with different BDWFs, computed analytically
vs. simulations. 
}
\label{tab:noise-comparison}
\end{table}

\section{Simulations and results}
\newcommand{\BIN}[2]{#1 s $\times$ #2 MHz}
In this section we use BDWFs to resample simulated visibility data, and study the effect on smearing and source 
suppression. Apart from a few examples documented separately, the basis interferometer configuration employed in 
the simulations corresponds to  \JVLA-C observing at 1.4 GHz. Similarly to Sect.~\ref{subsec:noise}, we create a 
``high-res'' measurement set corresponding to a 400 s synthesis at 1s integration, with 30 MHz total bandwidth
centred on 1.4 GHz, divided into 360 channels of 83.4 kHz each.
The MS is populated by noise-free simulated visibilities corresponding to a single point source at 
a given distance from the phase centre. We then generate ``low-res'' MSs to receive the resampled visibilities,
resample the latter using a range of BDWFs, convert the visibilities to dirty images (using natural weighting
unless otherwise stated), and measure the peak source flux in each image. Since each dirty image corresponds to 
a single source, the peak flux gives us \ATM{the degree of smearing or the smearing factor (i.e. the  amplitude decrease  for off-axis sources)
associated with a given BDWF and sampling interval.}

For the first set of simulations, the ``low-res'' MSs correspond\ATM{s} to a 100 s and 10 MHz synthesis. We employ three 
sampling rates, \BIN{25}{2.5}, \BIN{50}{5} and \BIN{100}{10} (thus 4 timeslots and 4 channels, 2 timeslots
and 2 channels, and single-channel snapshot).

\begin{figure*}
\includegraphics[width=.9\textwidth]{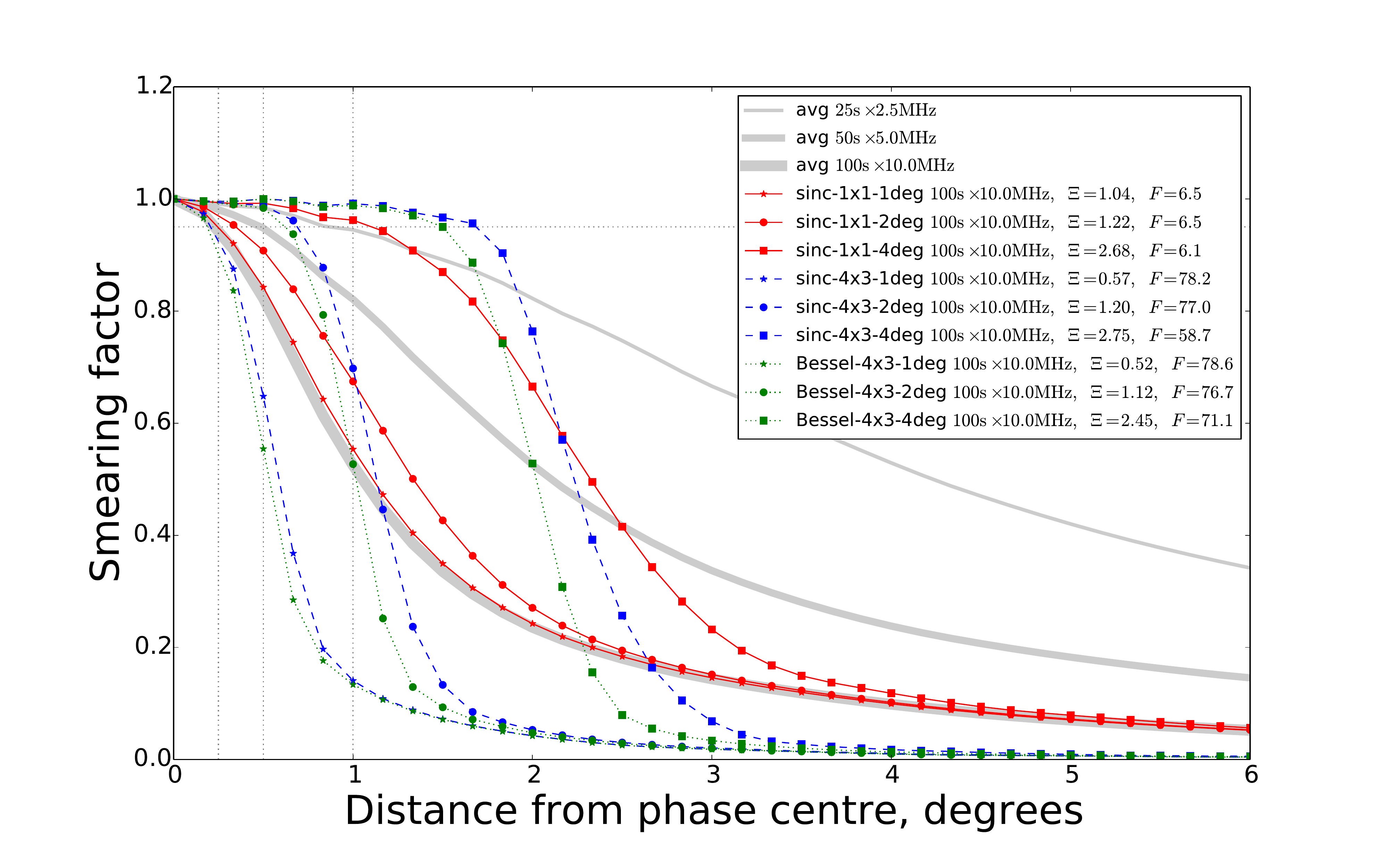}
\caption{
% \textcolor{red}{Reviewer comments: 1.10) Fig 8,9,10,11,12,13) I find these plots a bit difficult to use.
%            They seem similar and yet they have confusing differences, such as:
%            why is the y-axis labeled slightly different on 8 compared to rest,
%            the 3 variants of grey line plots have no visible differences
%            (whatever you do, don't torture yourself with too many shades of grey,
%            50 is enough :-). Maybe a few of the latter plots could be reduced to
%            a table with just the smear factor at the FoI edge or value at 95\% threshold.}
           \ATM{\JVLA-C observing at 1.4 GHz for the 400 s snapshot and 30 MHz band synthesis; illustrating  the performance of non-overlapping and overlapping 
BDWFs tuned to three FoIs for the same $uv$-bin size and the  associated noise penalty}. Smearing as a function of distance from \ATM{the} phase centre, for conventional averaging with 
\BIN{25}{2.5}, \BIN{50}{5} and \BIN{100}{10} bins, and for several BDWFs with \BIN{100}{10} bins.
The noise penalty $\Xi$ and the far-source suppression factor $F$ are given relative to \BIN{25}{2.5}
averaging.}
\label{fig:results-example}
\end{figure*}
\begin{figure*}
\includegraphics[width=.9\textwidth]{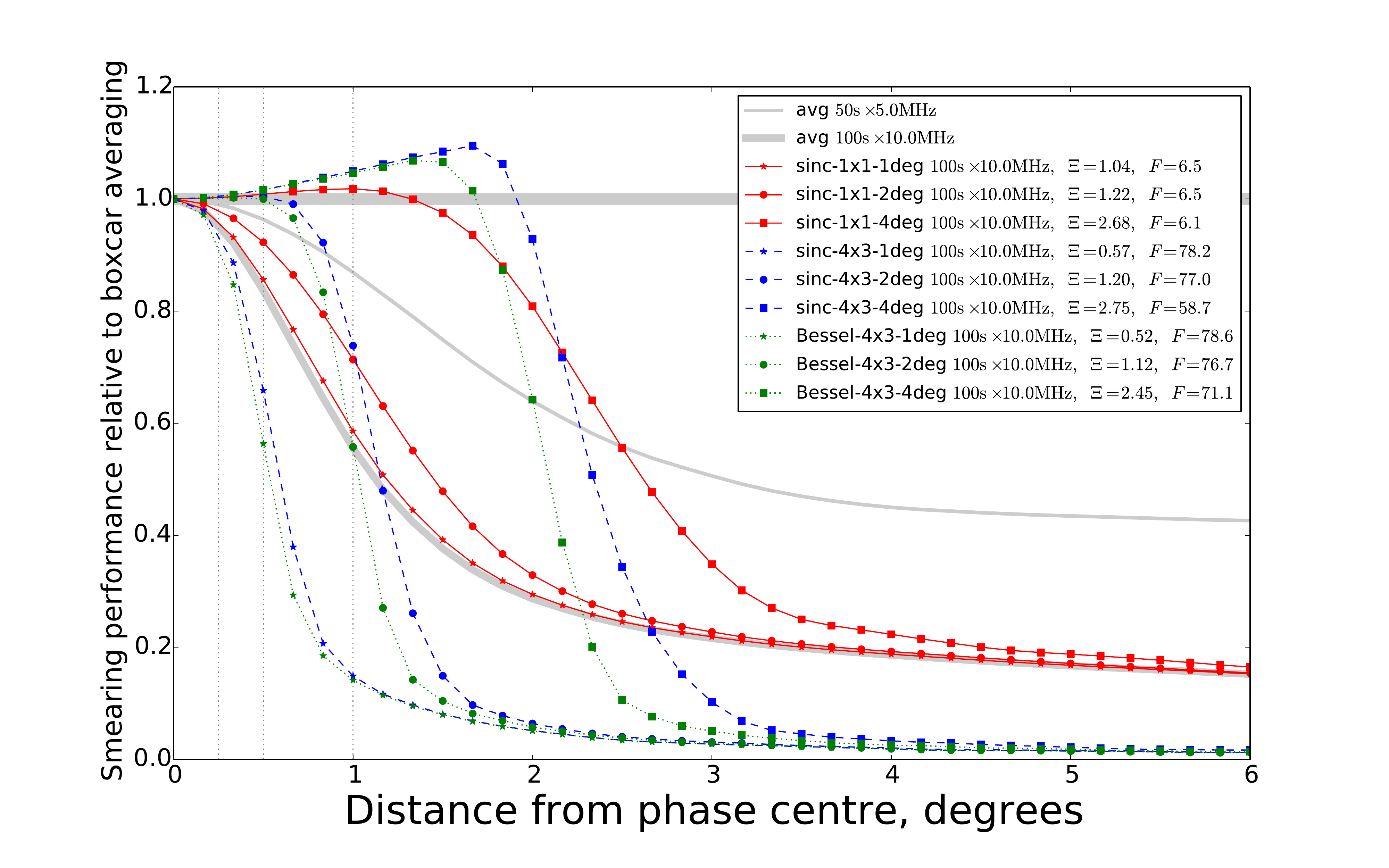}\\
\caption{
% \textcolor{red}{Reviewer comments: 1.10) Fig 8,9,10,11,12,13) I find these plots a bit difficult to use.
%            They seem similar and yet they have confusing differences, such as:
%            why is the y-axis labeled slightly different on 8 compared to rest,
%            the 3 variants of grey line plots have no visible differences
%            (whatever you do, don't torture yourself with too many shades of grey,
%            50 is enough :-). Maybe a few of the latter plots could be reduced to
%            a table with just the smear factor at the FoI edge or value at 95\% threshold.}
           \ATM{\JVLA-C observing at 1.4 GHz.} Results of Fig.~\ref{fig:results-example} normalised to the \BIN{25}{2.5} 
averaging curve, \ATM{the 
horizontal grey line at y=1 indicates that the \BIN{25}{2.5} averaging curve is normalised by itself.
This figure clearly illustrates the excellent performance of overlapping
BDWFs tuned to $2^\circ$.}}
\label{fig:results-example2}
\end{figure*}

A typical performance comparison for the \JVLA-C configuration at 1.4 GHz is given by Fig.~\ref{fig:results-example}.
This figure illustrates some of the principal achievements of the present work, so let us spend some time explaining it.
The horizontal axis represents distance from phase centre, while the vertical axis of the left-hand plot represents 
the smearing factor (left plot). Unity corresponds to no smearing; this is the case at phase centre, thus all curves
start at unity. The three thick gray curves correspond to normal averaging into \BIN{25}{2.5}, \BIN{50}{5} and
\BIN{100}{10}. We can (rather arbitrarily) define a series of ``acceptable'' smearing levels by specifying a  FoI
radius, and the maximum extent of smearing over that FoI. For the FoI radius, we may pick e.g. the half-power point  of
the primary beam, the main lobe of the primary beam, or extent of the first sidelobe of the primary beam. For \JVLA's 25 m dishes at this frequency,
these radii correspond to $\sim0.25^\circ$, $0.5^\circ$, and $1^\circ$, respectively; they are indicated by thin
vertical lines in the figure. The thin horizontal line indicates our chosen smearing threshold of $0.95$. In the right
plot,  all the curves are normalised with respect to the \BIN{25}{2.5} averaging curve.

For regular averaging, the three chosen bin sizes happen to roughly correspond to acceptable levels of smearing over the
three chosen FoI values. The other curves show the performance of a few different BDWFs, all at 100 s $\times$ 10 MHz
sampling. There are three types of BDWFs shown, indicated by line style (and colour, in the colour version of the plot):
\begin{itemize}
\item \WF{\Sincc}{1}{1}: a non-overlapping \Sincc~window (solid line, red)
\item \WF{\Sincc}{4}{3}: an overlapping \Sincc~window (dashed line, blue)
\item \WF{\Bessel}{4}{3}: an overlapping Bessel window (dotted line, green)
\end{itemize}
These are tuned to three different FoI settings, as indicated by the plot symbol: $1^\circ$ (star), $2^\circ$ (circle),
$4^\circ$ (square). 

The plot is meant to show performance of BDWFs at \BIN{100}{10} versus a ``baseline case'' of \BIN{25}{2.5} averaging, 
the latter being an acceptable averaging setting for this particular frequency and telescope geometry. The legend 
next to the plot therefore indicates $\Xi$, the noise penalty associated with that particular BDWF, and $F$, the 
far source suppression factor. Both values are calculated w.r.t. the baseline case. Note the following salient features:

\begin{itemize}
\item All overlapping BDWFs provide outstanding far source suppression in this regime, with $F$ in the $60\sim80$ range. 
The non-overlapping \Sincc~(solid red lines) only achieves $F~6$, which is similar to regular averaging
at the same rate. 
\item Noise performance is excellent for the $1^\circ$ BDWFs. There is a small noise penalty at $2^\circ$, and a larger 
(over a factor of 2) noise penalty at $4^\circ$. This can be easily understood by considering the shape of BDFWs as
a function of FoI: smaller FoIs correspond to broader windows that become more ``boxcar-like'' over the sampling
interval, and vice versa. This means that, in this particular configuration, BDWFs cannot achieve a FoI of $4^\circ$
at \BIN{10}{100} without a substantial sacrifice in sensitivity. We shall return to this issue below.
\item If the desired FoI size is $r\sim0.5-1^\circ$, overlapping BDWFs (\Sincc-4x3-2deg and \Bessel-
4x3-2deg) provide excellent performance at \BIN{100}{10}. Compared to averaging at \BIN{25}{2.5}, they achieve a factor
of 16 data compression with minimal loss of sensitivity, with excellent tapering behaviour: the smearing performance 
across the FoI is equivalent to (or better) than that of normal averaging, and out-of-FoI source suppression is almost
two orders of magnitude higher.
\end{itemize}

Fig.~\ref{fig:results-example2} presents the same results in an alternative way. Here, the recovered flux is shown 
relative to the baseline case of \BIN{25}{2.5} averaging. This clearly illustrates the excellent performance of overlapping
BDWFs tuned to $2^\circ$.

\subsection{Noise penalties and overlapping BDWFs}
Values of $\Xi<1$ above may be paradoxical at first, since one \textcolor{black}{can not} theoretically exceed the noise performance of the
unweighted average. This, however, is an artefact of our short simulation. Overlapping BDWFs are essentially
averaging in ``bonus signal'' from regions of overlap extending outside the nominal time and frequency 
coverage. In our case, at \BIN{100}{10} sampling, a BDWF with $4\times3$ overlap is actually adding up signal over a
\BIN{400}{30} bin, i.e. a bin that is a factor of 12 larger (though of course the bonus sensitivity thus gained is much
less than the theoretically available $\sqrt{12}$, since the weights over the overlap regions correspond
to the ``wings'' of the window function, and are thus small). This can easily result in lower per-visibility noise than that 
achieved by regular averaging over \BIN{100}{10}, and correspondingly higher snapshot sensitivity.

In the more realistic case of a long, multiple-channel synthesis (what \textcolor{black}{we will} call a \emph{full synthesis}), the
effects of bonus sensitivity disappear. While the noise on individual visibilities remains nominally lower in a full 
synthesis thanks to the overlap, it becomes correlated across neigh-boring $uv$-bins, so there is no net gain 
in image-plane sensitivity. Strictly speaking, at the ``edge'' of the synthesis, overlapping BDWFs are still pulling 
in some bonus signal from overlap regions extending beyond the synthesis coverage, but since the area of this 
overlap is negligible compared to the coverage of the full synthesis, so is the effect of the bonus signal.
\begin{figure*}
\includegraphics[width=.9\textwidth]{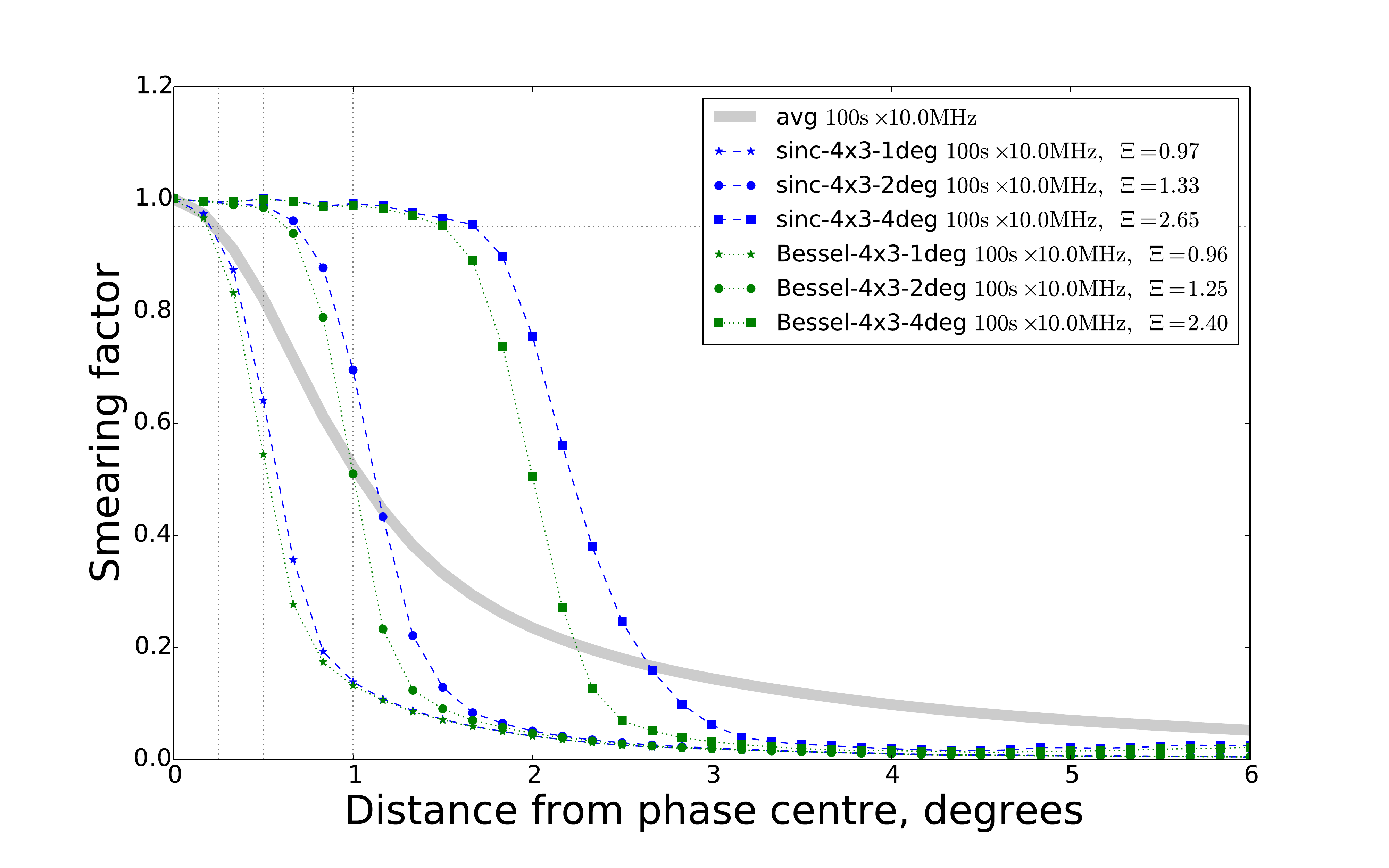}\\
\caption{
% \textcolor{red}{Reviewer comments: 1.10) Fig 8,9,10,11,12,13) I find these plots a bit difficult to use.
%            They seem similar and yet they have confusing differences, such as:
%            why is the y-axis labeled slightly different on 8 compared to rest,
%            the 3 variants of grey line plots have no visible differences
%            (whatever you do, don't torture yourself with too many shades of grey,
%            50 is enough :-). Maybe a few of the latter plots could be reduced to
%            a table with just the smear factor at the FoI edge or value at 95\% threshold.}
           \ATM{\JVLA-C observing at 1.4 GHz for the full 30 min and 200 MHz band synthesis; illustrating  the performance of overlapping 
BDWFs tuned to three FoIs for the same $uv$-bin size, and the associated  noise penalty.}
Smearing as a function of distance from \ATM{the} phase centre, 
for conventional averaging with \BIN{100}{10} bins, and for overlapping BDWFs with \BIN{100}{10} bins.
The noise penalty $\Xi$ is given relative to \BIN{100}{10} averaging.}
\label{fig:results-longsynth}
\end{figure*}

In other words, simulating a snapshot observation results in underestimated noise penalties, compared to the
real-life case of a full synthesis. We should expect the noise penalties to go up (and eventually exceed unity) as 
we increase the synthesis time and number of channels. Fig.~\ref{fig:results-longsynth} presents the results of such 
a simulation. This shows a a \BIN{1800}{200} synthesis, sampled at the same rates as above. The results should be 
compared to and contrasted with those of Fig.~\ref{fig:results-example}. Note that the tapering response of BDWFs is 
nearly identical, while the noise penalties are indeed higher. With $4\times3$ overlap and \BIN{100}{10} sampling, 
the total signal accessed by overlapping BDWFs corresponds to \BIN{2100}{220}, which gives a theoretical maximum of 
a factor of $\sim1.13$ in bonus sensitivity. In other words, the values of $\Xi$ in Fig.~\ref{fig:results-longsynth}
are still underestimated, but by 13\% at most (which explains $\Xi<1$ for the $1^\circ$ case). From this we may safely
extrapolate that the noise penalty of BDWFs matched to $1-2^\circ$ FoIs will remain reasonable even for a much longer 
and wider band synthesis.

\subsection{FoIs and sampling rates}
For BDWFs, a given FoI tuning represents a characteristic scale in the $uv$-plane, which is inversely proportional to the 
FoI parameter. On the other hand, the $uv$-bin sampled by any given visibility is proportional to the integration time,
fractional bandwidth, and baseline length. Since the window function is truncated at edge of the averaging bin (which can be
larger by the sampling bin by a factor of several, if overlapping BDWFs are employed), there is, for any given baseline,
some kind of optimal range of $uv$-bin sizes over which BDWFs tuned to a particular FoI setting are ``efficient''. Over 
smaller bins, BDWFs become equivalent to a boxcar averaging, over larger bins, BDWFs penalise too much
sensitivity as they down-weigh more samples. Since this optimal bin size is proportional to baseline length, the overall
optimum is dependent on the distribution of baselines in the array. 

Furthermore, the sampling rate needs to be ``balanced'' in time and frequency for BDWFs to achieve efficient tapering
response. If the $uv$-bins are elongated, the window function becomes truncated (i.e. more boxcar-like) across the bin, which reduces its ability 
to induce the desired taper. Since the orientation of the bins changes as the baseline rotates, the 
cumulative effect is an average degradation of the tapering response, in the
sense that it becomes closer to that of boxcar averaging. In this sense,
the optimal $uv$-bin shape is square-like. This occurs when the fractional bandwidth is equal to the arc section swept
out by the baseline over one bin. For a polar observation (circular $uv$-tracks), we can express this as
\begin{equation}
\Delta \nu/\nu = 2 \pi (B_x/B)(\Delta t/24\mathrm{h}),
\end{equation}
where $B$ is the baseline length, and $B_x$ is its East-West component. Rewriting this in terms of more convenient units,
we \ATM{obtain}
\begin{equation}
\frac{\Delta \nu_\mathrm{MHz}}{\Delta t_\mathrm{s}} \approx \frac{\nu_\mathrm{MHz}}{14000} \cdot \frac{B_x}{B},
\label{eq:optimal-binsize} 
\end{equation}
leading to a simple rule-of-thumb: at 1.4 GHz, an East-West baseline sweeps out a square-like bin when the 
integration time in seconds is 10 times the channel width in MHz (hence the use of bin sizes such as \BIN{100}{10} 
in the analysis here). 
\begin{figure*}
\includegraphics[width=.9\textwidth]{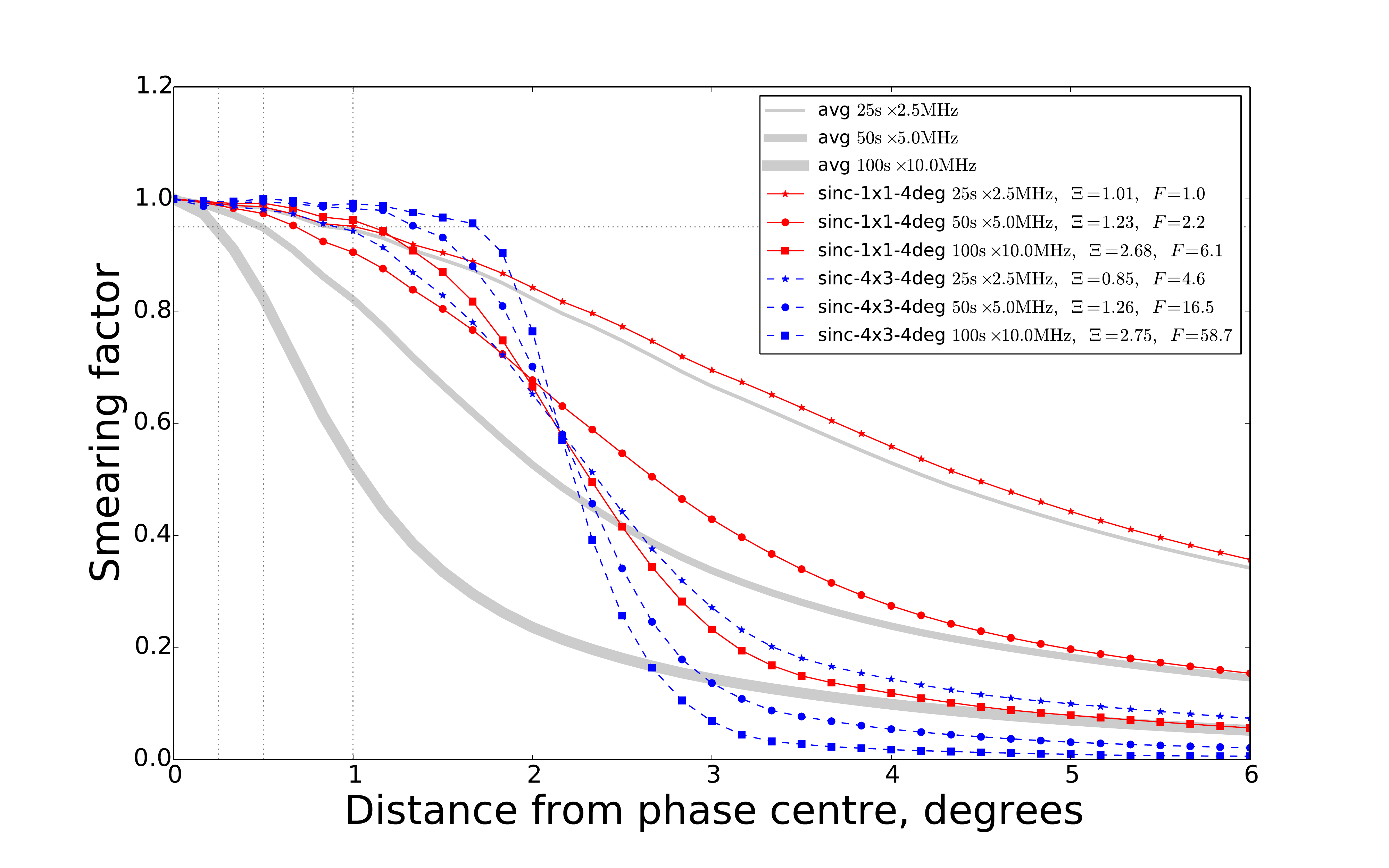}
\caption{
% \textcolor{red}{Reviewer comments: 1.10) Fig 8,9,10,11,12,13) I find these plots a bit difficult to use.
%            They seem similar and yet they have confusing differences, such as:
%            why is the y-axis labeled slightly different on 8 compared to rest,
%            the 3 variants of grey line plots have no visible differences
%            (whatever you do, don't torture yourself with too many shades of grey,
%            50 is enough :-). Maybe a few of the latter plots could be reduced to
%            a table with just the smear factor at the FoI edge or value at 95\% threshold.}
           \ATM{\JVLA-C observing at 1.4 GHz
for the 400 s snapshot and 30 MHz band synthesis; illustrating  the performances of non-overlapping and overlapping 
BDWFs tuned to 4$^{\circ}$ FoI and for three  $uv$-bin size, with the associated noise penalty.}
Smearing as a function of distance from \ATM{the} phase centre, for conventional averaging with 
\BIN{25}{2.5}, \BIN{50}{5} and \BIN{100}{10} bins, and for several BDWFs with \BIN{100}{10} bins.
The noise penalty $\Xi$ and the far-source suppression factor $F$ are given relative to \BIN{25}{2.5}
averaging.}
\label{fig:results-3sincs}
\end{figure*}
\begin{figure*}
\includegraphics[width=.9\textwidth]{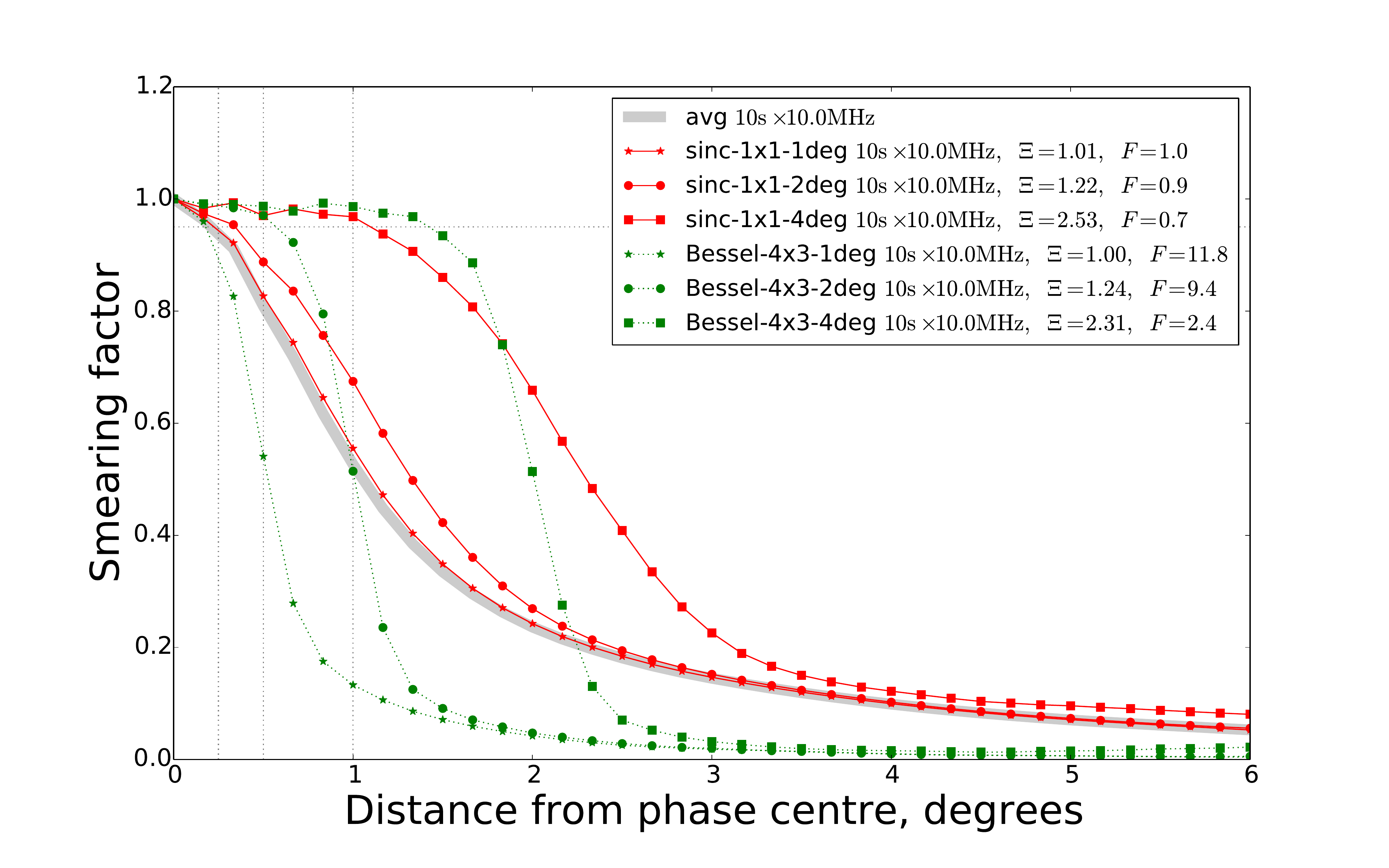}
\caption{
% \textcolor{red}{Reviewer comments: 1.10) Fig 8,9,10,11,12,13) I find these plots a bit difficult to use.
%            They seem similar and yet they have confusing differences, such as:
%            why is the y-axis labeled slightly different on 8 compared to rest,
%            the 3 variants of grey line plots have no visible differences
%            (whatever you do, don't torture yourself with too many shades of grey,
%            50 is enough :-). Maybe a few of the latter plots could be reduced to
%            a table with just the smear factor at the FoI edge or value at 95\% threshold.}
           \ATM{\JVLA-C observing at 14 GHz. This shows that the results of the BDWFs are completely determined by the $uv$-plane 
geometry in wavelengths. This figure is equivalent to \JVLA-C observing at a frequency of 1.4 GHz, with baselines scaled up by a factor of 10.}
Smearing as a function of distance from \ATM{the} phase centre, for conventional averaging with 
\BIN{2.5}{2.5}, \BIN{5}{5} and \BIN{10}{10} bins, and for several BDWFs with \BIN{10}{10} bins.
The noise penalty $\Xi$ and the far-source suppression factor $F$ are given relative to \BIN{2.5}{2.5}
averaging.}
\label{fig:results-14ghz}
\end{figure*}

The interaction between $uv$-bin size and tapering response is illustrated \ATM{in} Fig.~\ref{fig:results-3sincs}. Here we
compare the performance of two BDWFs tuned to a $4^\circ$ FoI -- a non-overlapping \WF{\Sincc}{1}{1} filter 
(solid red lines) and an overlapping \WF{\Sincc}{4}{3} filter (dashed blue lines) -- over three sampling bin sizes: 
\BIN{25}{2.5}, \BIN{50}{5} and \BIN{100}{10}. For reference, the performance of boxcar averaging over the same bin sizes
is indicated by the thick grey lines. Note how at the smaller bin size, the non-overlapping \Sincc~is practically 
equivalent to a boxcar in terms of tapering response; at the larger bin size, it begins to shape the FoI. Introducing an 
overlap improves the response considerably. An overlapping filter at \BIN{25}{2.5} achieves almost the same tapering
response as a non-overlapping one at \BIN{100}{10} (which is not surprising, if one considers that the effective 
averaging bin size in the former case is \BIN{100}{7.5}). However, for all filters, at \BIN{100}{10} the noise penalty 
goes up quite sharply. 

This illustrates that \BIN{50}{5} is an appropriate BDWF sampling rate for achieving a $4^\circ$ FoI (for \JVLA-C configuration at 
1.4 Ghz), providing a reasonable trade-off between tapering response and noise penalty. At higher sampling rates, 
the tapering response is degraded, while at lower sampling rates, the noise penalty increases. In comparison (as we 
saw in the previous section), for FoIs of $1-2^\circ$, BDWFs achieve a good trade-off at \BIN{100}{10} sampling.

It is interesting to consider how optimal BDWF sampling changes as a function of array size. Fig.~\ref{fig:results-14ghz}
shows a simulation for \JVLA-C at 14 GHz. (Since our results are completely determined by $uv$-plane geometry in 
wavelengths, this is equivalent to \JVLA-C scaled up by a factor of 10 at an observing frequency of 1.4 GHz). 
From eq.~\ref{eq:optimal-binsize}, we can see that square-like $uv$-bins correspond to 
sampling rate combinations such as \BIN{10}{10}. The simulation presented here is for a \BIN{1800}{200} synthesis,
i.e. is closer to the full synthesis rather than a snapshot case. Comparing Figs.~\ref{fig:results-14ghz} and 
\ref{fig:results-longsynth}, we find nearly identical BDWF performance (in terms of tapering response and
noise penalty) at 14 GHz and 1.4 GHz, with only the optimal sampling rate being different. 

\subsection{BDWFs for wide-field VLBI}
In the VLBI case, it is usually a combination of smearing and data rates, rather than the primary beam, 
that effectively limits the FoI. For example, the current European VLBI Network (EVN) correlator 
\citep{keipema2015sfxc} operated by the Joint Institute for VLBI in Europe (JIVE) is capable of
producing data at dump rates down to 10 ms, with 16 MHz of total bandwidth split into up to 8192 channels. 
The maximum available FoI of an EVN experiment is restricted  by the smallest primary beam, which is usually 
that of the 100 m Effelsberg telescope -- or about $10'$ in diameter at L-band. The ENV 
calculator\footnote{{\tt http://www.evlbi.org/cgi-bin/EVNcalc}} shows that a dump rate of 0.125 s and 1024 
channels (16 kHz) is required to keep smearing to within 10\% across this FoI. Due to the large computational 
and storage requirements, such data rates have only been employed in one-off experiments. For routine use, 
techniques such as  multiple-phase centre correlation are more common. Typically, data is averaged into more modest 
sampling rates of 2 s and 32 channels. This restricts the effective (L-band) FoI to about $20''$, and thus limits the scientific usefulness of archival data to narrow-FoI experiments.

\begin{figure*}
\includegraphics[width=.9\textwidth]{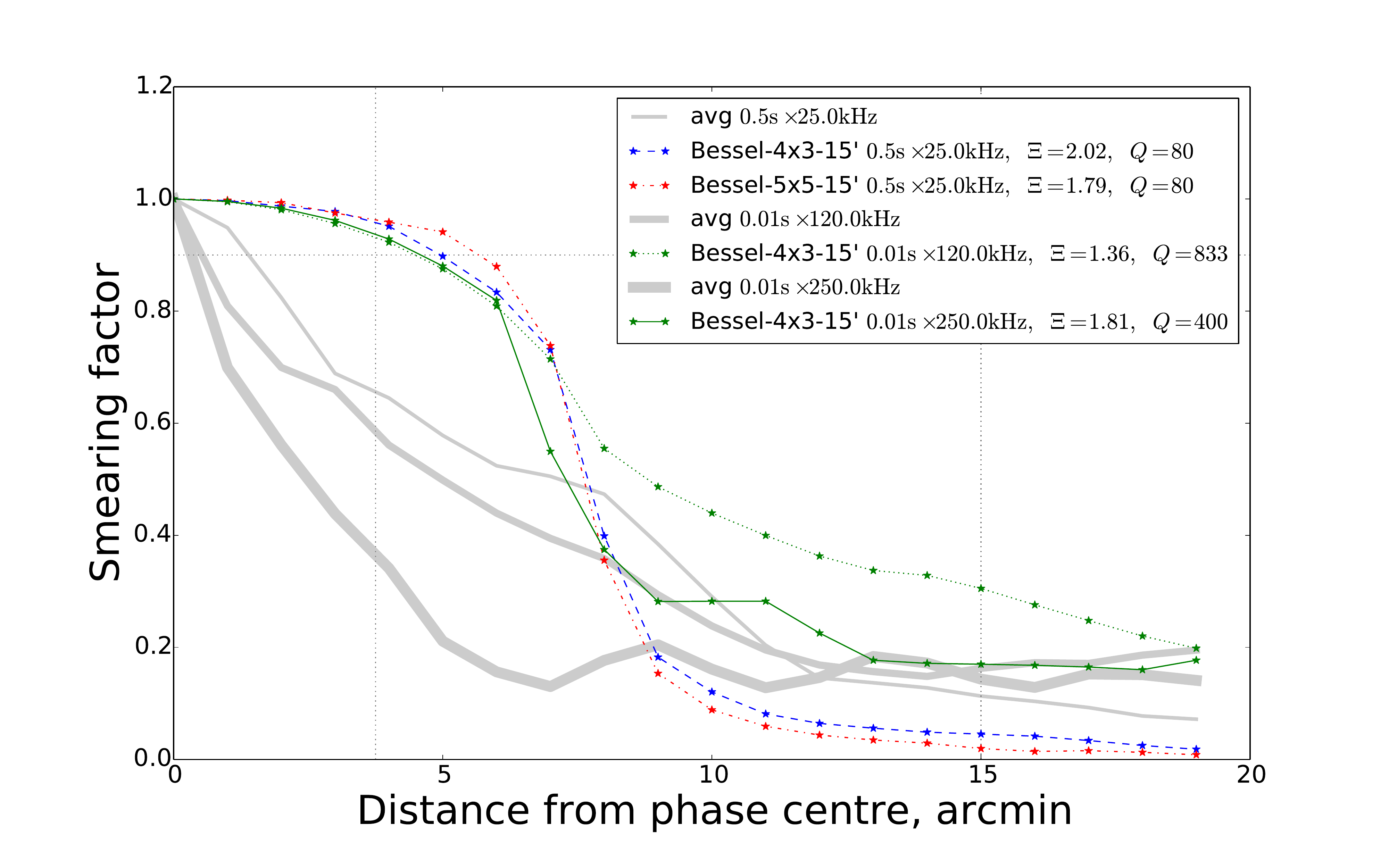}
\caption{VLBI simulation (1016 1km maximum baseline) at 1.6 GHz. Smearing as a function of distance from \ATM{the}
phase centre, for conventional averaging at 0.5 s $\times$ 25 kHz, 10 ms $\times$ 120 kHz and 10 ms $\times$ 250 kHz averaging, 
and overlapping Bessel BDWFs. The horizontal line corresponds to 10\% smearing losses. The ratio $Q$ shows the 
increase in data volume over conventional 2 s $\times$ 0.5 MHz averaging.
}
\label{fig:results-vlbi}
\end{figure*}

In this section we investigate whether the use of BDWFs can enable true wide-field VLBI. We simulate a 1.6 GHz 
EVN observation employing eight stations (Effelsberg, Hartebeesthoek, Jodrell Bank, Noto, Onsala, Torun, \JVLA, 
Westerbork, Shanghai), with a maximum baseline of 10161 km. Fig.~\ref{fig:results-vlbi} compares the smearing 
response of normal averaging to that of two overlapping Bessel BDWFs, employing 0.5 s and 25 kHz sampling. 
At these data rates, it becomes almost practical to have a full-FoI EVN archive. 

For comparison, we also show the performance of BDWFs for a hypothetical fast-transient archive application. In 
order to localise potential fast radio bursts (FRBs), we would need to retain a high time resolution of 10 ms, as 
well as a lower spectral resolution for de-dispersion. In this regime, BDWFs are less efficient since the 
$uv$-bins are elongated. Fig.~\ref{fig:results-vlbi} shows that this translates into less source suppression outside the FoI, but does not impact the ability to retain FoI.

\section{Conclusions and future work}
The goal of this work was to demonstrate the application of baseline-dependent window functions to radio 
interferometry. We have demonstrated that BDWFs offer a number of interesting advantages over conventional averaging.
The first of these is data compression -- i.e. visibilities can be sampled at a
lower rate, while retaining a large FoI. Compression by a factor of 16 with relatively little loss of sensitivity has been demonstrated. The huge data 
rates from upcoming instruments such as ASKAP, MeerKAT and the future SKA1 mean that raw visibilities may need to be
discarded after calibration (unlike older instruments, where raw visibilities have typically been archived). This 
represents something of a risk to the science, as it precludes future improvements in calibration techniques from 
being later applied to the data. With BDWFs, at least a highly-compressed version of the visibilities may be retained.

The second potential benefit of BDWFs is the increased suppression of unwanted signal from out-of-FoI sources. This
reduces both the overall level of far sidelobe confusion noise, and lessens the impact of A-team sources in sidelobes.

Thirdly, BDWFs can have an interesting impact in the VLBI case, as they allow the full primary beam FoI to be 
imaged using a single VLBI dataset. This opens the door to wide-field VLBI, which has previously been impractical.

BDWFs have a number of potential downsides. The first one is a potential loss in sensitivity. Our simulations show 
that this can be kept within reasonable limits, especially if overlapping BDWFs are employed, and can be traded off with
compression rate. Note that from DSP theory, we know that a properly matched filter can actually increase the SNR: the equivalent
result with BDWFs is that off-axis sources are attenuated less, so the effective SNR off-axis can increase despite the 
loss in absolute sensitivity.

The second downside of BDWFs is an increase in computational complexity. Whether implemented in a correlator
or in post-processing, BDWFs (and especially overlapping BDWFs) require substantially more operations than simple averaging.
There may be other limits to the practical applicability of BDWFs. They are far less efficient if high spectral  
resolution is required, so their use may be limited to continuum observations. Furthermore, averaging over longer 
intervals requires accurate phase calibration, so high compression rates may only be achievable post-calibration.

An interesting avenue of future research is combining BDWFs with baseline-dependent averaging. As we saw above, the ability
of BDWFs to shape a FoI is somewhat limited by the fact that shorter baselines sweep out smaller bins in $uv$-space, with 
window functions over them becoming boxcar-like. If baseline-dependent averaging is employed, shorter baselines are averaged over
larger $uv$-bins, thus increasing the effect of BDWFs. 

Finally, we should note that the use of BDWFs results in a different position-dependent PSF than regular averaging (or
to put it another way, the smearing response of BDWFs results in a different smeared PSF shape). Future work will focus
on methods of deriving this PSF shape, with a view to incorporating this into current imaging algorithms.
\section*{Acknowledgements}
This work is based upon research supported by the South African Research Chairs Initiative of the Department of 
Science and Technology and National Research Foundation.
The EVN-related research in this paper emerged from fruitful  discussions with Dr Aard Keimpema and Dr Zsolt Paragi at the 
Joint Institute for VLBI ERIC, whom we would like to thank for this collaboration. The visit to JIVE ERIC was made possible by the FP7 MIDPREP program.
We would like to thank the anonymous referee for comments that substantially improved the paper.
\bibliographystyle{mn2e}
\bibliography{m_paper}
\appendix
\section{Relative Performance of Window Functions}
\label{appendixA}
\newcommand{\FilterFigure}[4]{
\begin{figure}
\includegraphics[width=.5\columnwidth]{#1}%
\includegraphics[width=.5\columnwidth]{#2}
\caption{#3}\label{#4}
\end{figure}
}
Window functions -- or rather their corresponding image-plane response (IPR) -- can be characterised in terms of various metrics. Some common 
ones are the peak sidelobe level (PSL), the main lobe width (MLW) and the sidelobes roll-off (SLR) rate. In terms of the ``ideal'' 
IPR (Fig.~\ref{fig:idealwindowfunction}, left), these correspond to the following desirable traits:
\begin{itemize}
\item Maximally conserve the signal within the FoI (\textcolor{black}{Regime 1} in the figure),
and make the transition region (\textcolor{black}{Regime 2}) as sharp as possible. Both of these correspond to larger main lobe width.
\item Attenuate sources outside the FoI (\textcolor{black}{Regime 3}): this corresponds to a lower peak sidelobe level and higher sidelobes roll-off.
\end{itemize}
\begin{table}
\centering
\begin{tabular}{||l||ll|l|l||}
\hline
   \footnotesize window  &&{ \hspace{-0.8cm}\footnotesize $\sim$MLW} & { \footnotesize PSL}  & {\footnotesize  SLR}   \\
  \footnotesize functions &&\hspace{-0.8cm}(\footnotesize deg and at -3dB) & \footnotesize (dB) & \footnotesize (dB/oct)  \\
\hline\hline
{\footnotesize $\Pi(t/t_a)$}  & $t\in|t_a|$& $ 1.406$ &$-6.663$ &$-12.089$\\
	    & $t\in|t_a/2|$&$ 2.812$ &$-6.671$ &$-11.065$\\
\hline
{$\Sincc(t)$} & $t\in|t_a|$ &$ 12.304$& $-10.889$&  $-12.661$ \\
	 & $t\in|t_a/2|$ &$ 12.304$& $-13.241$&  $-11.447$ \\
\hline
{$J_0(t)$}& $t\in|t_a|$ &$ 9.140$ &$ -14.553$ & $ -12.011$\\
	  & $t\in|t_a/2|$ &$ 9.140$ &$ -13.614$ & $ -11.794$\\
\hline
{$\mathrm{G}(t)$} & b=3 &$ 2.109$& $-21.535$& $-9.589$\\ 
	 & b=5 &$ 2.812$& $-30.211$& $-9.091$\\ 
\hline
{$\mathrm{BW}(t)$} & p=1 &$ 2.109$ &$-13.718$ & $-12.581$\\
	  & p=3 &$ 4.218$ &$-10.145$ & $-27.330$\\
\hline
{$\mathrm{PS}_0(t)$} & $\alpha=2\pi^3$ &$ 3.515$& $-45.302$& $-7.424$\\ 
	 & $\alpha=5\pi^3$ &$ 4.218$& $-73.597$& $-6.375$\\ 
\hline
{$J_0^\mathrm{Hm}(t)$} & $t\in|t_a|$ &$ 9.140$&$-35.724$ & $-11.948$\\ 
 & $t\in|t_a/2|$ &$ 9.140$&$-22.670$&$-19.527$\\ 
\hline
{$\Sincc^\mathrm{Hm}(t)$}  & $t\in|t_a|$ &$ 12.656 $&$-27.581$ &$-13.817$ \\ 
 &$t\in|t_a/2|$ &$12.656$&$-13.469$&$-14.324$\\
 \hline
\end{tabular}
\caption{\label{tab:WF:performance}Comparative performance of 
different window functions. MLW: main lobe width,  PSL: peak sidelobe level, SLR: sidelobes roll-off. 
% \textcolor{red}{ Reviewer comments Sections:
%      1.1) Section 3 was confusing. The title suggests making a connection
%           to DSP theory which is not that important to have it's own section. It
%           then turns out that this is where you generalize the standard one
%           dimensional tapers to 2D tapers, am I correct?
%           Suggest removing DSP parts and moving the 2D taper generalizations to the
%           definitions of $X_pqkl$, eq 36, in terms of \Sincc~and Bessel functions.
%           (You really need explicit definitions of the BDWF you use.) Also try
%           to motivate why you chose these and not other functions.
%      1.2) The appendix, which is 3 pages long, does not provide more than
%           a basic summary of windowing/taper functions. Suggest removing it completely.
%           (Especially as you do not even use most of these functions in the main text.)
%           It suffices to have a reference to a signal processing text, or may be save just
%           the Table.}
          Here $\Pi$: boxcar window, $\mathrm{J}_0$: Bessel function of the first kind of order 0~\citep{watson1995treatise}, $\mathrm{G}$: Gaussian window, $\mathrm{BW}$: Butterworth window, $\mathrm{PS}_0$: prolate spheroidal wave function of sequence zero, $\Sincc^\mathrm{Hm}(t)$: \Sincc~multiplied by the Hamming function, $J_0^\mathrm{Hm}(t)$: Bessel multiplied by the Hamming function.}
\end{table}
Table~\ref{tab:WF:performance} summarises the performance of the different window functions. 
This table shows that the \Sincc, the Bessel and all their derivative with Hamming, Han and Blackman window functions 
have large main lobe,  low peak sidelobe level and high sidelobes roll-off and therefore provide the more optimal tapering response. 
\section{Overlapping BDWFs}
\label{appendixB}
\begin{figure}%
\centering
\includegraphics[width=.4\textwidth]{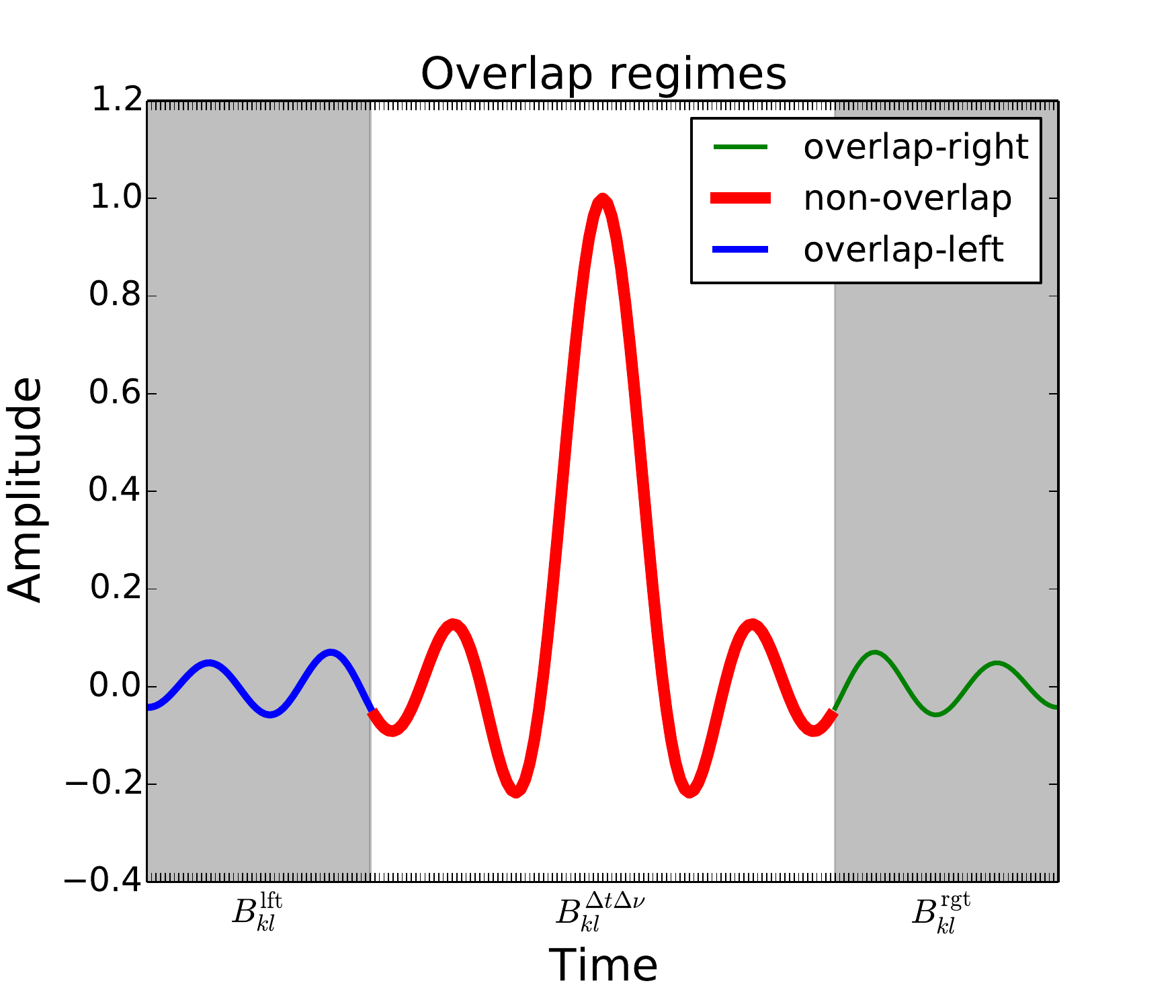}\\
\includegraphics[width=.4\textwidth]{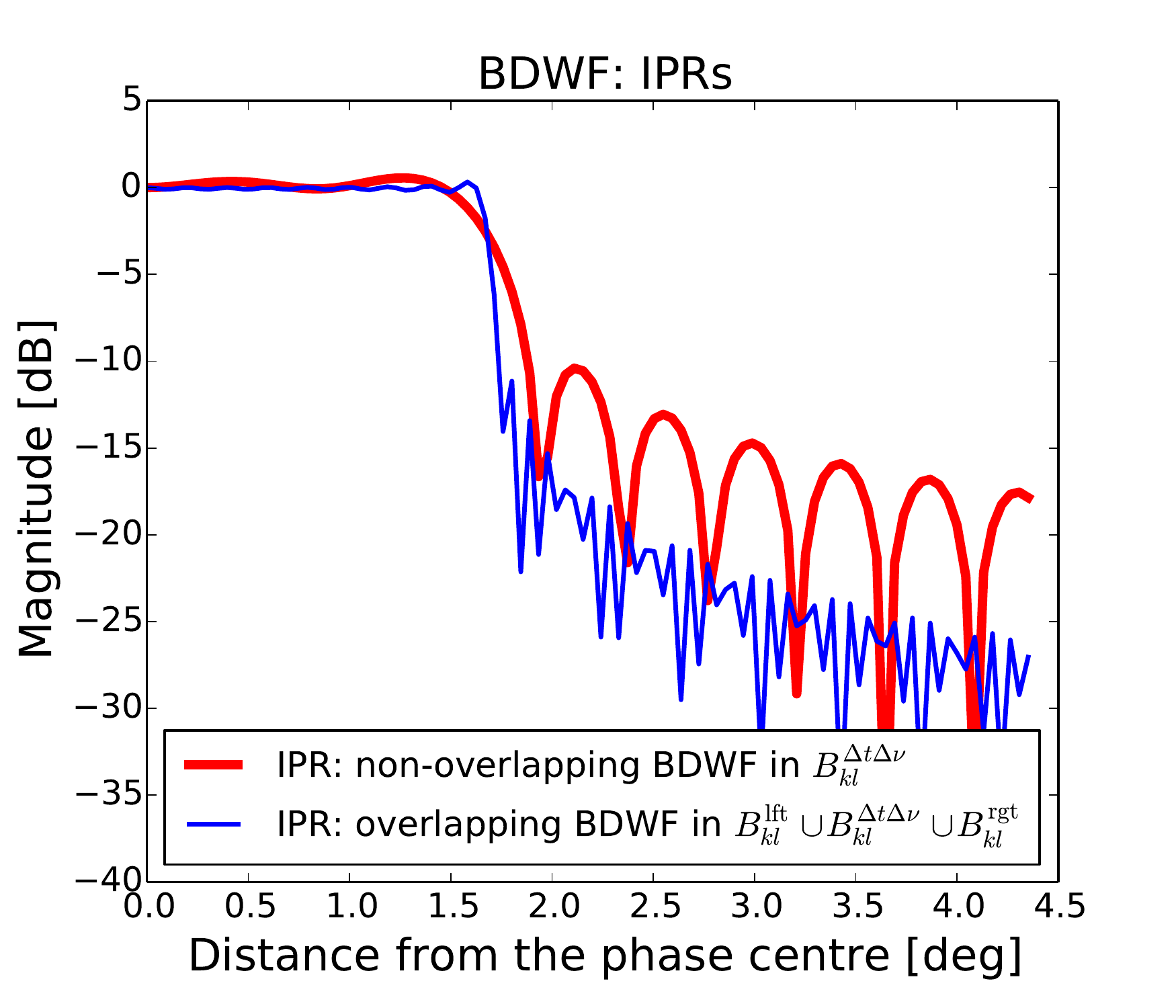}%
\caption{ \ATMNEW{Overlapping BDWF. (Top) Blue curve: left-hand overlap function, green curve: right-hand  overlap function, red curve: non-overlap function. (Bottom) The IPRs of an overllaping (blue) and non-overlapping (red) BDFW.}
}\label{fig:overlap-regime}
\end{figure}

\newcommand{\Btfleft}{\mathsf{B}^{\mathrm{lft}}}
\newcommand{\Btfright}{\mathsf{B}^{\mathrm{rgt}}}
\newcommand{\BOVERLAP}{\mathsf{B}^{[\Delta^\mathrm{olp} t,\Delta^\mathrm{olp}\nu]}}
Let us reconsider that $X$ is a BDWF with its appropriate resampling bin $\Btf_{kl}$.
Now, suppose that  $\Delta^\mathrm{lft} t$ and $\Delta^\mathrm{lft} \nu$ are 
the overlap time-frequency sampling intervals associated
with  $\Delta t$ and $\Delta \nu$ on their \textit{left-hand side}. Similarly, $\Delta^\mathrm{rgt} t$ and 
$\Delta^\mathrm{rgt} \nu$ are the overlap time-frequency sampling intervals associated
with  $\Delta t$ and $\Delta \nu$ on their \textit{right-hand side}. The overlapping 
BDWFs sampling intervals are  then given by: 
\begin{alignat}{2}
 \Delta^\mathrm{olp} t&=\Delta^\mathrm{lft} t \cup\Delta t \cup\Delta^\mathrm{rgt}t \\
 \Delta^\mathrm{olp} \nu &=\Delta^\mathrm{lft} \nu \cup \Delta\nu \cup \Delta^\mathrm{rgt} \nu.
\end{alignat}
The overlap sampling bin is now defined as: 
\begin{alignat}{2}
\BOVERLAP_{kl} &= \bigg [ t_k-\frac{\Delta^\mathrm{olp} t}{2},t_k+\frac{\Delta^\mathrm{olp} t}{2} \bigg ]\nonumber\\
&\times \bigg [ \nu_l-\frac{\Delta^\mathrm{olp}\nu}{2},\nu_l+\frac{\Delta^\mathrm{olp}\nu}{2} \bigg ]\\
	      &=\Btfleft_{kl}\cup\Btf_{lk}\cup\Btfright_{kl},
\end{alignat}
where $\Btfleft_{kl}$ and $\Btfright_{kl}$ are the set of time-frequency samples in the overlap regimes, and   $\Btf_{kl}$ is the resampling bin for a non-overlapping BDWFs defined in eq.~(\ref{eq:chap3resamplingbin}). Fig.~\ref{fig:overlap-regime} (top) displays the time direction overlap regimes (i.e. $\Btfleft_{kl}$ and $\Btfright_{kl}$).
We note that the  window bin defined in eq.~(\ref{eq:windowbinalphabeta}) for overlap factors of $\alpha,\beta$ is equivalent to
\begin{equation}
\Bab_{kl} = \big \{ ij:~t_i\nu_j \in \BOVERLAP_{kl} \big \}
\end{equation}
or in  more detail we can write:
\begin{alignat}{2}
\Bab_{kl} &= \Big \{ ij:~t_i\nu_j \in \Btfleft_{kl} \Big \} \cup \Big \{ ij:~t_i\nu_j \in \Btf_{lk} \Big \}\nonumber\\
&\cup \Big \{ ij:~t_i\nu_j \in \Btfright_{kl} \Big \} \\
	  &=S_1 \cup \Bij_{kl} \cup S_2,
\end{alignat}
where $S_1=\big \{ ij:~t_i\nu_j \in \Btfleft_{kl} \big \}$ and $S_2=\big \{ ij:~t_i\nu_j \in \Btfright_{kl} \big \}$ are the number 
of time-frequency samples in the left-hand and right-hand overlap regimes respectively. 
The  resampled visibilities for overlapping BDWFs (compare to eq.~(\ref{eq:avg:wf})) are as follows:
\begin{alignat}{2}
\Vm_{pqkl} &= \frac{\displaystyle\sum\limits_{{i,j}\in \Bab_{kl}} \Vs_{pqij} X(\bmath{u}_{pqij}-\bmath{u}_{pqkl})}
{\displaystyle\sum\limits_{{i,j}\in \Bab_{kl}} X(\bmath{u}_{pqij}-\bmath{u}_{pqkl})}.\label{eq:overlapvisibilities}
\end{alignat}
One may still decompose eq.~(\ref{eq:overlapvisibilities}) as:
\begin{alignat}{2}
\Vm_{pqkl} &= \frac{Q^\mathrm{lft}+
\displaystyle\sum\limits_{{i,j}\in \Bij_{kl}} \Vs_{pqij} X(\bmath{u}_{pqij}-\bmath{u}_{pqkl})+Q^\mathrm{rgt}
}
{R^\mathrm{lft}+\displaystyle\sum\limits_{{i,j}\in \Bij_{kl}} X(\bmath{u}_{pqij}-\bmath{u}_{pqkl})+R^\mathrm{rgt}},\label{eq:1overlapvisibilities}
\end{alignat}
where:\\
$Q^\mathrm{lft}=\displaystyle\sum\limits_{{i,j}\in S_1} \Vs_{pqij} X(\bmath{u}_{pqij}-\bmath{u}_{pqkl})$, \\
$Q^\mathrm{rgt}=\displaystyle\sum\limits_{{i,j}\in S_2} \Vs_{pqij} X(\bmath{u}_{pqij}-\bmath{u}_{pqkl}$,\\
$R^\mathrm{lft}=\displaystyle\sum\limits_{{i,j}\in S_1} X(\bmath{u}_{pqij}-\bmath{u}_{pqkl})$,\\
$R^\mathrm{rgt}=\displaystyle\sum\limits_{{i,j}\in S_2} X(\bmath{u}_{pqij}-\bmath{u}_{pqkl})$.\\
% \begin{alignat}{2}
% \Vm_{pqkl} &= \frac{\displaystyle\sum\limits_{{i,j}\in S_1} \Vs_{pqij} X(\bmath{u}_{pqij}-\bmath{u}_{pqkl})+
% \displaystyle\sum\limits_{{i,j}\in \Bij_{kl}} \Vs_{pqij} X(\bmath{u}_{pqij}-\bmath{u}_{pqkl})+
% \displaystyle\sum\limits_{{i,j}\in S_2} \Vs_{pqij} X(\bmath{u}_{pqij}-\bmath{u}_{pqkl})
% }
% {\displaystyle\sum\limits_{{i,j}\in S_1} X(\bmath{u}_{pqij}-\bmath{u}_{pqkl})+\displaystyle\sum\limits_{{i,j}\in \Bij_{kl}} X(\bmath{u}_{pqij}-\bmath{u}_{pqkl})+
% \displaystyle\sum\limits_{{i,j}\in S_2} X(\bmath{u}_{pqij}-\bmath{u}_{pqkl})}.\label{eq:1overlapvisibilities}
% \end{alignat}
Fig.~\ref{fig:overlap-regime} shows an overlapping BDWF $X(\bmath{u}_{pqij}-\bmath{u}_{pqkl})$ with ${ij}\in S_1$ (or $t_i\nu_i \in \Btfleft_{kl}$), 
$X(\bmath{u}_{pqij}-\bmath{u}_{pqkl})$ with ${ij}\in\Bij_{kl}$  
(or $t_i\nu_i \in \Btf_{lk}$) and $X(\bmath{u}_{pqij}-\bmath{u}_{pqkl})$ with $ {ij}\in S_2$ (or $t_i\nu_j \in \Btfright_{lk}$).
We note that eq.~(\ref{eq:1overlapvisibilities}) becomes equivalent to the non-overlapping resampled 
visibilities in eq.~(\ref{eq:avg:wf}) when $S_1=S_2= \emptyset$. This is 
due to the sum over an empty set $\emptyset$: \\
$Q^\mathrm{lft}=Q^\mathrm{rgt}=R^\mathrm{lft}=R^\mathrm{rgt}=0$.
% \begin{eqnarray}
%  \left\{ 
%   \begin{array}{l l}
%    Q^\mathrm{lft}=0, \\
% Q^\mathrm{rgt}=0,\\
% R^\mathrm{lft}=0,\\
% R^\mathrm{rgt}=0.\\
%   \end{array} \right.
% \end{eqnarray}
%\bsp
\label{lastpage}
\end{document}